\documentclass[10pt,conference]{IEEEtran}
\IEEEoverridecommandlockouts

\usepackage{pifont}
\usepackage{etoolbox}
\usepackage{amsmath,amsfonts}
\usepackage{amsthm}

\usepackage{graphicx}
\usepackage{cite}
\usepackage[ruled,vlined]{algorithm2e}
\usepackage{amsmath}
\usepackage{hyperref}
\usepackage[utf8]{inputenc}
\usepackage{tcolorbox}
\usepackage{lipsum}
\usepackage{multirow}
\usepackage{enumitem}
\usepackage{tabularx}
\usepackage{booktabs}  
\usepackage{listings} 
\usepackage{makecell}
\usepackage{graphicx}
\usepackage{subcaption}
\usepackage{booktabs}  
\usepackage{xcolor}    
\usepackage{colortbl}  
\usepackage{seqsplit}
\usepackage{tikz}
\usepackage[table]{xcolor}
\usepackage{multirow}
\usepackage{booktabs}
\definecolor{oursblue}{RGB}{226,240,253}
\definecolor{headgray}{RGB}{224,224,224}
\definecolor{wingreen}{RGB}{20,120,60}
\definecolor{failred}{RGB}{200,40,40}
\newcommand{\fw}[1]{\textbf{\textcolor{wingreen}{#1}}}
\newcommand{\fail}[1]{\textcolor{failred}{\textbf{#1}}}
\newcommand{\ob}{\cellcolor{oursblue}}

\definecolor{CodeGreen}{rgb}{0.13,0.55,0.13}  
\definecolor{CodeGray}{rgb}{0.5,0.5,0.5}      
\definecolor{CodePurple}{rgb}{0.58,0,0.82}    
\definecolor{CodeBlue}{rgb}{0.0,0.2,0.6}      
\definecolor{BackColour}{rgb}{0.97,0.97,0.97} 

\AtEndPreamble{
	\usepackage{hyperref}

	\hypersetup{
		colorlinks = true,
		linkcolor = purple,
		anchorcolor = purple,
		citecolor = purple,
		filecolor = purple,
		urlcolor = purple
	}
}

\lstdefinestyle{model}{
    language=,                                  
    basicstyle=\ttfamily\scriptsize,
    backgroundcolor=\color{BackColour},
    frame=l,
    rulecolor=\color{CodeGray!40},
    framesep=3pt,
    xleftmargin=10pt,
    xrightmargin=5pt,
    numbers=none,
    breaklines=true,
    showstringspaces=false,
    keywords={entity,use,released,by,repairs,found,obligation,verdict,reasoning,after,release,verdict,reasoning},
    keywordstyle=\bfseries\color{CodeBlue},
    keywords=[2]{VULNERABLE,UNDISCHARGED},
    keywordstyle=[2]\bfseries\color{CodePurple},
    keywords=[3]{SAFE,DISCHARGED,none},
    keywordstyle=[3]\bfseries\color{CodeGreen},
    morecomment=[l]{//},
    commentstyle=\itshape\color{CodeGray},
    sensitive=true,
    keepspaces=true,
}

\usepackage{listings}
\usepackage{xcolor}

\definecolor{listing-bg}{gray}{0.96}      
\definecolor{listing-frame}{gray}{0.8}    
\definecolor{listing-key}{RGB}{20, 60, 110} 

\lstdefinestyle{verdict-box}{
    backgroundcolor=\color{listing-bg},
    frame=single,
    framerule=0.4pt,
    rulecolor=\color{listing-frame},
    basicstyle=\ttfamily\scriptsize,
    keywordstyle=\color{listing-key}\bfseries,
    morekeywords={verdict, reasoning},
    breaklines=true,
    breakatwhitespace=true,
    breakindent=2em,       
    columns=fullflexible,
    keepspaces=true,
    xleftmargin=0.8em,
    xrightmargin=0.8em,
    aboveskip=0.8\baselineskip,
    belowskip=0.6\baselineskip,
    captionpos=t,           
    numbers=none,           
}

\lstdefinestyle{json}{
    language=,                                  
    basicstyle=\ttfamily\scriptsize,
    backgroundcolor=\color{BackColour},
    frame=single,
    rulecolor=\color{CodeGray!40},
    framesep=4pt,
    framerule=0.4pt,
    xleftmargin=8pt,
    xrightmargin=5pt,
    numbers=none,
    breaklines=true,
    breakatwhitespace=true,
    showstringspaces=false,
    stringstyle=\color{CodePurple},
    string=[s]{"}{"},
    morecomment=[l]{//},
    commentstyle=\itshape\color{CodeGray},
    literate=
        *{:}{{{\color{CodeGray}{:}}}}{1}
         {,}{{{\color{CodeGray}{,}}}}{1}
         {\{}{{{\color{CodeBlue}\{}}}{1}
         {\}}{{{\color{CodeBlue}\}}}}{1}
         {[}{{{\color{CodeBlue}[}}}{1}
         {]}{{{\color{CodeBlue}]}}}{1},
    keywords=[2]{VULNERABLE,UNDISCHARGED,Invalidation,Usage},
    keywordstyle=[2]\bfseries\color{CodePurple},
    keywords=[3]{SAFE,DISCHARGED,none},
    keywordstyle=[3]\bfseries\color{CodeGreen},
    sensitive=true,
    keepspaces=true,
    columns=fullflexible,
}

\newcommand{\tool}{\textsc{LeanGuard}}

\begin{document}

\title{Neuro-Symbolic Reasoning for Vulnerability Detection}

\author{
\IEEEauthorblockN{
Yanjie Zhao\textsuperscript{1},
Hongjie Chen\textsuperscript{1},
Li Lu\textsuperscript{1},
Zhou Yang\textsuperscript{2},
Xiao Cheng\textsuperscript{3},
and Haoyu Wang\textsuperscript{1}\ding{41}
}
\IEEEauthorblockA{\textsuperscript{1}Huazhong University of Science and Technology, Wuhan, China\\
\{yanjie\_zhao, hongjie\_chen, li\_lu, haoyuwang\}@hust.edu.cn}
\IEEEauthorblockA{\textsuperscript{2}University of Alberta; Alberta Machine Intelligence Institute; Canada CIFAR AI Chair, Canada, zy25@ualberta.ca}
\IEEEauthorblockA{\textsuperscript{3}Macquarie University, Sydney, Australia, xiao.cheng@mq.edu.au}
\thanks{\ding{41} Corresponding author: Haoyu Wang (haoyuwang@hust.edu.cn).}
}

\maketitle

\begin{abstract}
Ask a large language model (LLM) whether a pointer dereference is safe, and it can often produce a plausible justification for ``yes''. The difficulty is that a fluent justification is not a proof. This gap is precisely where automated vulnerability detection lives: deciding, for a given operation in source code, whether a memory safety defect such as a null dereference, use-after-free, or double free can actually occur. We trace the unreliability of LLM-based vulnerability detection to a mechanism, the \emph{premature discharge} of safety obligations, and argue that the remedy is not better prompting but a separation of roles: the component that interprets the code must not also be the one that decides a safety obligation is met.

In this paper, we present \tool{}, a neuro-symbolic framework that assigns each act to the side equipped for it. On the neural side, an LLM serves strictly as a \emph{semantic filter} over candidate facts extracted from the abstract syntax tree (AST): it prunes spurious facts and keeps the real ones, but never discharges an obligation or decides the verdict on its own. On the symbolic side, the surviving facts are compiled into a verification model in \emph{Lean~4} (a formal proof assistant whose kernel accepts a conclusion only when it is formally proved), where every dangerous operation must be matched by a guard that provably covers it in scope; absent such a guard, the obligation stays open rather than being argued away. Because a function rarely arrives with full context, this symbolic model is necessarily partial: an unproved obligation is not yet a defect. An evidence-aware adjudicator therefore weighs the symbolic and neural verdicts by the quality of each. We instantiate the framework on five CWE classes to ask how far this division of labor can be pushed. Across these classes and three backends, including agentic baselines granted full-repository access, \tool{} improves F1 over the strongest baseline in all fifteen settings. On a repeated-run subset (three CWE classes $\times$ three backends), chosen to bound experimental cost, the gain is significant in six of nine ($p<0.05$). It helps most precisely where premature discharge hurts most: on CWE-415, recall roughly doubles, from 0.21 to 0.41.
\end{abstract}


\section{Introduction}
\label{sec:intro}

Memory safety defects remain among the most damaging and persistent flaws in systems software. Classes such as NULL pointer dereference (CWE-476), use-after-free (CWE-416), and double free (CWE-415) sit atop the most-reported weakness categories year after year~\cite{mitre_cwe_top25,mitre_cve_metrics}, and they survive even in mature, heavily audited code: the Linux kernel alone absorbs hundreds of NULL-dereference fixes in a single recent year, across memory management, file systems, and the network stack~\cite{nvd_linux_cwe476,cvedetails_linux_2024}. What makes them so stubborn is an asymmetry inherent to the task. A defender must guard \emph{every} dangerous operation; an attacker needs only the one that slipped through. Since these operations are scattered across a codebase that no analysis can hold in full, each must be judged where it sits---often without the rest of the program in view. The hard question is therefore no longer whether these bugs exist, but whether we can judge a single function reliably when complete context is \textit{not guaranteed} to be at hand.

Two families of techniques dominate automated vulnerability detection, and their limitations are complementary. Rule-driven static analyzers such as Infer~\cite{infer}, Coverity~\cite{coverity}, and Cppcheck~\cite{marjamaki2026cppcheck} are deterministic and auditable, but their reach is bounded by hand-curated rule bases that encode patterns humans have already distilled and carry no understanding of code semantics~\cite{DBLP:journals/cacm/BesseyBCCFHHKME10}. Large language models (LLMs) have the opposite profile: pretrained on vast corpora, they read code semantics and perform cross-function reasoning~\cite{10.1145/3695988,10.1145/3769676}, and recent agentic systems extend this toward autonomous security research~\cite{google_bigsleep,sakana_fugu,anthropic_mythos}. An LLM, however, is a probabilistic generator rather than a logical reasoner. It introduces variables that do not exist, misidentifies the direction of data flow, and commits to a verdict on insufficient evidence. 
Consistent with this, Deletang et al.~\cite{DBLP:conf/iclr/DeletangRGGWCCH23} show that Transformers fail to generalize on tasks requiring precise counting and recursive nesting, the very kind of structural bookkeeping that tracking obligations across control flow demands.
The verdict is also unauditable: when an LLM reports a function as safe, a genuine analysis of the code cannot be distinguished from a coincidental match to training data.

We locate the deeper failure not in the noise of LLM reasoning but in how that reasoning meets a proof obligation. Memory safety can be cast as \emph{obligation discharge}: every dereference, access, and free imposes a condition that must be shown to hold. An LLM treats such an obligation like any other generation target and discharges it as soon as it can phrase a fluent reason. \textit{\textbf{But a justification is not a discharge.}} A model that explains why a pointer is \emph{probably} non-null has not established that it \emph{is}. This gap closes with neither \textit{scale} nor \textit{search}. When we hand state-of-the-art coding agents such as Codex~\cite{openai2021codexweb} and Claude Code~\cite{anthropic2026claude-code} the entire repository and let them retrieve, on their own initiative, whatever context they judge relevant, they still systematically err toward declaring code safe, leaving genuine defects unflagged, most acutely on the lifecycle classes where safety hinges on temporal ordering (\autoref{sec:rq1}). Stronger models and more context buy more fluent justifications, not more discharged obligations; the extra reasoning only produces more plausible grounds for releasing an obligation that should have stayed open. The LLM should not be the component that discharges obligations.

This reframes the design question. Rather than make the LLM reason more reliably or the rules more flexible, we separate \emph{judging what the evidence means} from \emph{discharging an obligation}, and assign each to the side suited for it, while a structural extraction layer bounds what either side may assert in the first place. The structural layer grounds the analysis: it confines the neural model to editing a fixed list of syntactically extracted candidates rather than inventing protections from source, which is what keeps the model from flagging nearly everything as unsafe. Over these grounded facts, the neural model handles ambiguity: it decides which candidates are semantically real and which are spurious, without ever ruling on safety. The symbolic side then handles what the neural model is structurally ill-equipped for: it pairs each obligation with the protections that could cover it and holds the obligation open unless one structurally does. It refuses to release an obligation that nothing protects, and it is this structural discharge (faithful pairing followed by refusal, not neural fluency) that recovers the defects an LLM would have let pass. The LLM curates and interprets evidence; it never discharges an obligation, and it cannot close one by argument alone.

We realize this principle in \tool{}, which links an LLM semantic filter to a structural protection checker built on \textit{Lean~4}~\cite{10.1007/978-3-030-79876-5_37}, a dependent-type theorem prover, through a deterministic bridge. The LLM (whether a pure model or an agentic backend) is demoted from decision maker to fact editor, returning a constrained add-or-remove schema over a numbered AST candidate list rather than free-form prose, while a scope-aware pairing algorithm reconstructs control-flow protection relations and assembles the surviving facts into a valid Lean~4 model, so no ill-formed input reaches the checker. Crucially, \tool{} operates at the target-function level with no whole-program access: context is admitted on demand, and any obligation it cannot resolve locally is read conservatively rather than as proof of safety. Because the model is necessarily partial, no component holds sole authority; an evidence-aware adjudicator fuses the checker's machine-checked record of discharged and open obligations with the constrained neural verdict, weighting each by quality. We instantiate this pipeline on five representative CWE classes; the per-class specification it currently requires is an explicit boundary of the study.

This paper makes the following contributions:

\begin{itemize}[leftmargin=1em]
\item We identify the \emph{premature discharge} of safety obligations as the core reliability failure of LLM-based detection, and argue that proposing facts and discharging obligations must be separated so that no component holds sole authority.
\item We realize this separation in \tool{}: an instruction-based fact-editing schema and a scope-aware guard-to-operation pairing algorithm turn unconstrained code understanding into mechanically checkable input, which a declarative Lean~4 protection semantics verifies and an evidence-aware adjudicator fuses with a constrained neural verdict, reading unproved obligations conservatively.
\item Across five CWE classes and three backends (a pure LLM and two agents with full-repository access), \tool{} improves F1 over the strongest baseline in all fifteen settings, by up to 0.20 absolute, with the largest gains on lifecycle weaknesses such as CWE-415, where F1 rises from 0.34 to 0.54 and recall roughly doubles (0.21 to 0.41). 
\end{itemize}

\section{Background and Related Work}
\label{sec:background}

\subsection{Rule-Driven Static Analysis}
Declarative query languages such as CodeQL~\cite{github2026codeql} let analysts describe vulnerability patterns as logical rules; Infer~\cite{infer} applies separation logic to infer pre- and post-conditions across functions; and Coverity~\cite{coverity} and Cppcheck~\cite{marjamaki2026cppcheck} combine data-flow analysis, taint propagation, and pattern matching to flag common CWE classes. Their limitations are equally well understood: rule bases are expensive to maintain and must be re-adapted as code style or project structure changes, and limited semantic modeling makes implicit cross-function constraints hard to express. A pointer checked in a caller but dereferenced along another path, or a boundary condition that holds only inside a specific error-handling branch, routinely inflates false positive rates~\cite{staticfp1, staticfp2}.

\subsection{LLMs for Code Analysis}
Pretrained code models~\cite{codebert} gave machines vectorized representations of code semantics, and large-scale generators such as DeepSeek-Coder~\cite{deepseekcoder} advanced code understanding and generation. In security specifically, IRIS~\cite{iris} uses an LLM to generate taint specifications for a static analyzer, QLPro~\cite{qlpro} couples LLMs with CodeQL through role specialization, and fact-aligned, template-constrained generation~\cite{factalign} improves rule accuracy. All of these, however, operate inside a probabilistic inference frame in which the model output serves directly as the rule source or the final conclusion, with no independent verification barrier. In a security setting, where hallucination is dangerous, this motivates our choice to decouple probabilistic generation from strict logical verification.

\subsection{Formal Verification and Lean 4}
Theorem provers grounded in type theory give reasoning a mathematical level of rigor: once a safety property is proved, it holds on all execution paths the model admits. Landmark results such as the seL4 microkernel~\cite{sel4} and the CompCert compiler~\cite{compcert} show that formal methods, though costly, are decisive for high-assurance systems. Lean~4~\cite{10.1007/978-3-030-79876-5_37} doubles as a programming language and a proof system, with strong metaprogramming support~\cite{leanmeta} and a built-in decision procedure that discharges propositions over finite Boolean domains, lowering the barrier to automated verification. A growing body of work couples LLMs with provers~\cite{DBLP:journals/corr/abs-2509-22908,DBLP:conf/acl/XinXYCWXSZD25,DBLP:journals/corr/abs-2504-21801}, but largely targets proof-ability evaluation or automated proving itself; turning real code, with its complex control flow, implicit state, and engineering context, into verifiable propositions still incurs a high modeling cost.

\subsection{Neuro-Symbolic Systems}
A line of work closes the loop between LLMs and provers. APOLLO~\cite{apollo} pioneered a generate-verify-repair cycle in which the model proposes a proof, Lean returns feedback, and the model revises; DeepSeek-Prover~\cite{deepseekprover} adds reinforcement learning and Monte Carlo tree search, using verifier feedback as a policy signal. Cobbe et al.~\cite{cobbe} argue that an independent verifier substantially improves reasoning reliability, the role Lean~4 plays here. The decisive difference is one of intent: prior neuro-symbolic provers place the LLM on the generative critical path, asking it to produce proofs while the symbolic engine scores or refines them. \tool{} inverts this. Rather than translating code into proofs, the neural contribution is confined to bounded fact confirmation and constrained semantic review, while formal reasoning operates over a separately constructed model, sidestepping the error-prone code-to-proof translation and restricting the LLM to the localized judgment it performs reliably (\autoref{sec:approach}).

\section{Motivating Example}
\label{sec:motivating}

We adopt use-after-free (UAF) vulnerabilities (CWE-416~\cite{mitre2026cwe416}) as a running example throughout the paper. The specific case is CVE-2021-3640, a UAF in the Linux kernel Bluetooth SCO function \texttt{\seqsplit{sco\_sock\_sendmsg}}, whose relevant control flow is shown in \autoref{lst:vuln}.

\begin{lstlisting}[language=C,caption={Unpatched control flow of \texttt{sco\_sock\_sendmsg}.},label={lst:vuln}]
lock_sock(sk);
if (sk->sk_state == BT_CONNECTED)
    err = sco_send_frame(sk, msg, len);  // [1] Callee may invalidate sk
else
    err = -ENOTCONN;
release_sock(sk);  // [2] Later use of sk
return err;
\end{lstlisting}

The defect does not lie in \texttt{release\_sock}, which merely releases the socket lock. It arises from an implicit inter-procedural state change: \texttt{sco\_send\_frame} copies a user-supplied message while holding the lock, on a path that can invalidate the object reachable through \texttt{sk}. Formally, the execution path introduces a sequence $\langle \text{Invalidate}(sk), \text{Use}(sk) \rangle$ without an intervening $\text{Repair}(sk)$ or a terminal $\text{Abort}()$. 




This example is revealing because every neural baseline misses it. Across three independent runs each, the pure LLM (DeepSeek-V4-Pro~\cite{deepseekai2026deepseekv4}), the Codex agent~\cite{openai2021codexweb}, and the Claude Code agent~\cite{anthropic2026claude-code}, with both agents also configured to use the DeepSeek-V4-Pro API as their underlying model, each return \texttt{SAFE} with essentially the same justification: they erroneously assume unlocking a mutex implies the pointer is valid, missing the cross-call invalidation entirely (detailed in \autoref{lst:stage2_interaction}). Because the argument is informal, the same omission recurs across runs and across both the pure and agentic settings.

To resolve this, the upstream fix confirms the vulnerability by restructuring the data flow. As shown in \autoref{lst:patched}, the patch allocates a buffer and copies the user message before acquiring the lock, passing only the safe buffer to \texttt{sco\_send\_frame}.

\begin{lstlisting}[language=C,caption={Patched control flow of \texttt{sco\_sock\_sendmsg}.},label={lst:patched}]
buf = kmalloc(len, GFP_KERNEL);
/* ... copy msg into buf ... */
lock_sock(sk);
if (sk->sk_state == BT_CONNECTED)
    err = sco_send_frame(sk, buf, len, msg->msg_flags); // sk is safely decoupled
else
    err = -ENOTCONN;
release_sock(sk);
kfree(buf);
return err;
\end{lstlisting}

The patched version successfully isolates the risky sequence, breaking the data dependency between the object invalidation and the subsequent lock release. A dependable detector should be able to formally capture the risk in \autoref{lst:vuln} and correctly verify the safety of \autoref{lst:patched}. It should not discard a dangerous operation grounded in the source merely because an informal argument dismisses it. It should produce the same decision on repeated runs, and it should expose the evidence behind its verdict, so that the decision can be audited rather than trusted blindly. These strict requirements directly motivate the design of \tool{}.

\begin{figure*}[htbp]
    \centering
    \includegraphics[width=\linewidth]{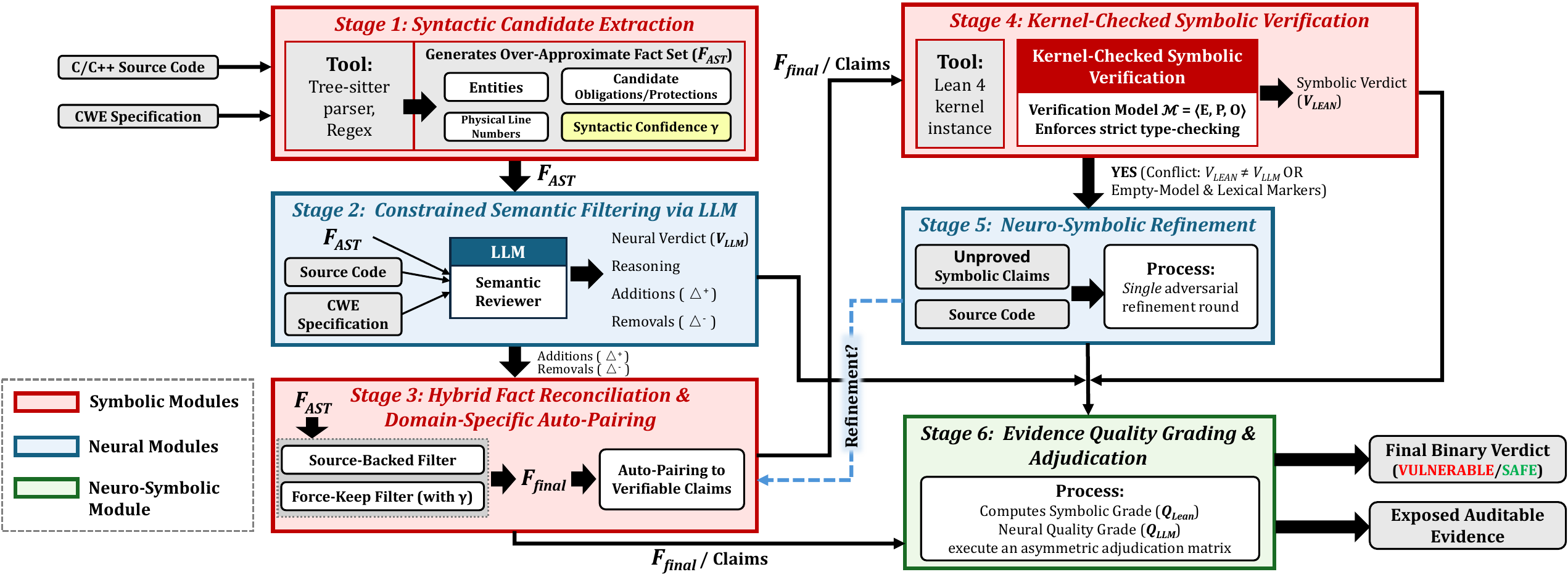}
    \caption{Overview of the \tool{} framework.}
    \label{fig:overview}
\end{figure*}

\section{Approach}
\label{sec:approach}

\tool{} aims to detect specific CWE vulnerabilities~\cite{mitre2024cwe} in C/C++ functions. Our approach relies on a neuro-symbolic architecture that pipelines six stages (\autoref{fig:overview}): syntactic candidate extraction, constrained neural semantic filtering, deterministic fact reconciliation and auto-pairing, kernel-checked symbolic verification, neuro-symbolic refinement, and evidence-aware adjudication.
A foundational principle of \tool{} is the strict separation of authority: the neural network acts solely as a semantic reviewer, while the symbolic verifier operates exclusively on synthesized logical models. \textit{In realistic settings, the analyzer rarely has access to the full program}: a function under review may invoke external callees, rely on macros expanded elsewhere, or manipulate objects whose allocation and release span other files. Because such context is often missing, \tool{} avoids whole-program safety certification and instead discharges localized structural obligations, weighting the resulting evidence by quality so that an incomplete symbolic model is interpreted conservatively rather than treated as ground truth.

\subsection{A Unified Formal Abstraction for Multi-CWE Semantics}
\label{sec:unified}

To handle structurally disparate software weaknesses under a single verification engine, \tool{} expresses each CWE class in a common abstraction. The abstraction is shared, but it is instantiated separately for one target class at a time: given a program and a fixed target CWE, \tool{} abstracts the program into a structured Verification Model $\mathcal{M} = \langle E, P, O \rangle$ whose components are populated according to that class's specification.
\begin{itemize}[leftmargin=1em]
    \item $E$ is the set of \emph{Entities}, the components whose safety-relevant state is tracked (e.g., pointer variables, allocated buffer objects). Every protection and obligation is bound to one such entity through an evaluation function $\mathrm{ent}(\cdot)$.
    \item $P$ is the set of \emph{Protections}, the explicit program operations or state guards that establish safety properties (e.g., nullity validation, boundary tracking, lifetime re-assignment).
    \item $O$ is the set of \emph{Obligations}. Each obligation $o \in O$ is a safety-critical operation that constitutes a potential vulnerability if executed without a corresponding protection.
\end{itemize}

In the implementation, each obligation is materialized together with its candidate protections as a single \emph{\textbf{claim}}, the atomic unit on which verification operates. Whereas an obligation $o$ denotes the abstract dangerous operation, a claim $c = \langle o, P_o \rangle$ is its concrete, verifiable encoding, pairing $o$ with the candidate protection set $P_o = \{\, p \in P \mid \mathrm{ent}(p) = \mathrm{ent}(o) \,\wedge\, \mathrm{kind}(p) \vdash_{CWE} \mathrm{kind}(o) \,\}$ of protections that share its entity \emph{and} are type-compatible, and could therefore discharge it. Here $\mathrm{kind}(x) \in \mathcal{T}_{CWE}$ is a type function ($\mathcal{T}_{CWE}$ denotes the operation types defined for each weakness class) and $\vdash_{CWE}$ is a domain-specific entailment over protection and obligation types; protections that fail either test (e.g., a dereference rendered unreachable by an intervening \texttt{return}) are filtered out and never enter $P_o$. An obligation $o$ is then discharged iff some protection in its (already type-eligible) set structurally covers it ($\prec_{\mathrm{safe}}$):
\begin{equation}
\label{eq:discharge_formal}
\mathrm{discharged}(o) \iff \exists\, p \in P_o.\ \bigl(p \prec_{\mathrm{safe}} o\bigr)
\end{equation}
The abstract dominance relation $\prec_{\mathrm{safe}}$ is instantiated per class: region containment for spatial weaknesses (the protection encloses the memory extent touched by $o$) and control-flow dominance for lifecycle weaknesses (the protection executes before $o$), whose obligations additionally carry a triggering release encoded as a dedicated \texttt{release} field alongside their \texttt{checks}. In practice, $\prec_{\mathrm{safe}}$ is resolved during Stage~3 auto-pairing wherever syntactically decidable (e.g., control-flow dominance, scope enclosure), so the Lean kernel discharges each claim over an already-paired model rather than re-deriving control flow; an obligation with $\{p \in P_o \mid p \prec_{\mathrm{safe}} o\} = \emptyset$ is reported as undischarged. This single relational interface, realized by class-specific implementations, lets heterogeneous weaknesses share one verification kernel. The model yields $V_{Lean} = \mathtt{VULNERABLE}$ whenever any obligation remains undischarged. Crucially, $V_{Lean}$ is not the final label: it is one structured input to the evidence-aware adjudicator, since an undischarged obligation may reflect extraction noise rather than a genuine defect (\autoref{sec:stage6}).

Beyond this relational polymorphism, generality also depends on the nature of the obligations themselves. We distinguish two regimes. For spatial and dereference weaknesses (CWE-476, CWE-120, CWE-125), an obligation is \emph{stateless}: the dangerous operation, such as a dereference, write, or read, is itself the unit to be discharged. For lifecycle weaknesses (CWE-416, CWE-415), an obligation is \emph{stateful}: it is meaningful only relative to a preceding lifecycle event, so its generation requires first identifying the triggering release. This is why the latter two classes track temporal ordering during extraction, while the former three reason purely over co-located structure.

To demonstrate the generality of this abstraction, we formally define the concrete mappings for the five supported CWE types\footnote{We instantiate these five representative CWE categories to demonstrate the feasibility of the approach; \textbf{the underlying methodology is inherently extensible to other memory safety and structural weaknesses}.}, summarized in \autoref{tab:cwe_mapping} and elaborated below.

\begin{table*}[t]
\centering
\caption{Instantiation of the unified Verification Model $\mathcal{M} = \langle E, P, O \rangle$ across the five supported CWE classes.}
\label{tab:cwe_mapping}
\resizebox{\linewidth}{!}{
\begin{tabular}{@{}lp{0.13\linewidth}p{0.2\linewidth}p{0.2\linewidth}p{0.45\linewidth}@{}}
\toprule
\textbf{CWE} & \textbf{Entities $E$} & \textbf{Protections $P$} & \textbf{Obligations $O$} & \textbf{Entailment $\vdash_{CWE}$ \& Coverage $\prec_{\mathrm{safe}}$} \\
\midrule
476 (Null Ptr. Deref.) & Pointer variables & Nullity assertions, conditional guards (\texttt{if (p)}) & Pointer dereferences (\texttt{*p}, \texttt{p->f}) & Any well-formed nullity guard discharges any dereference (single protection type). \\
\midrule
120 (Buffer Overflow) & Bounded buffer objects & Spatial boundary checks (\texttt{sizeof(buf)}, loop limits) & Memory write sequences (\texttt{memcpy}, \texttt{strcpy}) & Name-level match between checked boundary and destination buffer. \\
\midrule
125 (Out-of-Bounds Read) & Readable memory regions & Index-like or capacity-like spatial checks & Structural memory reads & Index-like reads require an index check; streaming reads require a capacity check. \\
\midrule
416 (Use-After-Free) & Object allocation lifecycles & Lifetime repairs (\texttt{p = NULL}, reacquisition, reallocation) & Accesses within a post-release window & A repair discharges an access only if it dominates the access on every path. \\
\midrule
415 (Double Free) & Allocated objects & State resets after the first release, or control-flow exits preventing re-entry & Release operations subsequent to the first release & A later release is discharged only if a valid intervening state reset dominates it. \\
\bottomrule
\end{tabular}
}
\end{table*}

\begin{itemize}[leftmargin=1em]
    \item \textbf{CWE-476 (Null Pointer Dereference):} $E$ is the set of pointer variables. $P$ consists of nullity assertions and conditional guards (e.g., \texttt{if (p)}). $O$ is the set of pointer dereference operations (e.g., \texttt{p->field}, \texttt{*p}). The relation $\vdash_{476}$ holds for any well-formed nullity guard, since null-safety admits a single protection type and thus imposes no further type discrimination.
    \item \textbf{CWE-120 (Classic Buffer Overflow):} $E$ represents bounded buffer objects. $P$ comprises syntactic spatial boundary checks (e.g., \texttt{sizeof(buf)}, loop limits). $O$ models memory write sequences (e.g., \texttt{memcpy}, \texttt{strcpy}). The relation $\vdash_{120}$ performs name-level matching between check boundaries and destination buffers.
    \item \textbf{CWE-125 (Out-of-Bounds Read):} $E$ represents readable memory regions. $P$ comprises spatial constraints categorized into index-like or capacity-like checks. $O$ models structural memory reads. The relation $\vdash_{125}$ strictly dictates that an index-like read requires an index check, whereas a parser-based streaming read requires a capacity check.
    \item \textbf{CWE-416 (Use-After-Free):} $E$ tracks object allocation lifecycles. $P$ models lifetime repair events (e.g., pointer nullification \texttt{p = NULL}, reacquisition, reallocation). $O$ models accesses occurring within a post-release window, so each obligation is anchored to a distinct preceding release event identified during extraction.
    \item \textbf{CWE-415 (Double Free):} $E$ tracks allocated objects. The first release of an object establishes its lifecycle state and is not itself an obligation; only a \emph{subsequent} release constitutes an obligation in $O$. $P$ models state resets following the first release (e.g., pointer nullification) or control-flow exits that prevent re-entry. The relation $\vdash_{415}$ dictates that a subsequent release is discharged only if a valid intervening state reset strictly dominates it.
\end{itemize}

\subsection{Stage 1: Syntactic Candidate Extraction}
\label{sec:stage1}

The initial stage processes the raw source text to construct an over-approximate, high-recall set of candidate program facts derived from the Abstract Syntax Tree (AST), denoted $F_{AST}$. Recognizing that false negatives at this juncture are unrecoverable downstream, the extraction engine prioritizes structural recall over localized precision.

The pipeline uses tree-sitter~\cite{tree-sitter} parsing paired with localized regular-expression compensation to handle the irregularities of partial code blocks and macros. Each extracted candidate fact is tracked as a multi-dimensional record containing its functional classification (e.g., Guard, Operation, Release), the target entity identifier, its physical line number, and the literal code expression.

The extractor emits no confidence annotation. Instead, once $F_{AST}$ is materialized, a post-extraction pass scans the recovered facts against a set of CWE-specific risk lexicons and assigns each a syntactic confidence score $\gamma \in \{\mathrm{High}, \mathrm{Low}\}$. These lexicons enumerate operations and markers empirically prone to participating in the target weakness: for example, unbounded copy primitives such as \texttt{strcpy} and \texttt{sprintf} for CWE-120, and, for the lifecycle classes, a curated set of release and invalidation signals (e.g., \texttt{free}/\texttt{kfree} family deallocators, refcount drops such as \texttt{kref\_put} and \texttt{sock\_put}, reallocation-induced invalidations like \texttt{realloc} and \texttt{pskb\_expand\_head}) together with lifetime-repair signals (e.g., reference reacquisition, ownership transfer, nullification, and control-flow exits). For brevity, we report only these representative categories here; the complete pattern lexicons are provided in our artifact repository. A fact matching such a pattern is tagged $\gamma = \mathrm{High}$; all others default to $\gamma = \mathrm{Low}$. The tag carries no semantic verdict; it acts as a protective anchor for the Force-Keep filter of Stage~3, preventing the neural reviewer from discarding structurally salient sinks and repair evidence. This tagged fact set forms the output of Stage~1.

For our running example (\autoref{lst:vuln}), Stage 1 extracts \texttt{sk} as the target entity. Crucially, it captures the cross-call invalidation risk and the subsequent usage, mapping them to precise physical boundaries. The abstracted fact set $F_{AST}$ is formalized as shown in \autoref{lst:stage1_extract}.

\begin{lstlisting}[
    style=json,
    caption={Abstracted syntactic candidate facts ($F_{AST}$) extracted in Stage 1 for the running example.},
    label={lst:stage1_extract}
]
{
  "Target_Entity": "sk",
  "Candidate_Obligations": [
    { "type": "Invalidation", "location": "line 3: sco_send_frame(sk, msg, len)" },
    { "type": "Usage", "location": "line 6: release_sock(sk)" }
  ],
  "Candidate_Protections": []
}
\end{lstlisting}

\begin{lstlisting}[
    style=json,
    caption={Abridged neuro-semantic filtering interaction, illustrating a \textbf{neural false negative} for the running example.},
    label={lst:stage2_interaction}
]
// --- SYSTEM PROMPT (Abstracted) ---
Task: Filter structural UAF candidates. Do NOT dismiss cross-call invalidations.
Input:
 - Invalidation event at Line 3: sco_send_frame(...)
 - Usage event at Line 6: release_sock(sk)
Instruction: Output JSON with `remove_uses` if Line 6 is strictly safe.

// --- LLM RESPONSE (Semantic Failure) ---
{
  "verdict": "SAFE",
  "reasoning": "release_sock() at Line 6 merely releases a mutex and does not free memory. The pointer 'sk' remains valid.",
  "remove_uses": [ "line 6" ]
}
\end{lstlisting}

\subsection{Stage 2: Constrained Semantic Filtering via LLM}
\label{sec:stage2}

Because the syntactic candidate set naturally contains over-approximate noise, Stage 2 routes the raw source code, the specific CWE specification, and the enumerated candidate lists into an LLM tasked with serving strictly as a structured semantic reviewer.

To prevent the model from deviating into narrative dialogue or synthesizing hypothetical patches, it is tightly constrained by an immutable JSON output schema. The LLM must output a binary programmatic verdict ($V_{LLM}$), a brief reasoning string, and explicit lists of requested additions ($\Delta^+$) and removals ($\Delta^-$) over guards, operations, and tracked entities.

The instructions force the model to evaluate context-sensitive contracts that elude rigid syntax trees. However, the model often falls into semantic traps. In our running example, the LLM outputs $V_{LLM} = \mathtt{SAFE}$ and places the \texttt{release\_sock} operation into its $\Delta^-$ removal queue, erroneously reasoning that a mutex unlock inherently implies pointer safety and thereby dismissing the preceding cross-call invalidation risk (\autoref{lst:stage2_interaction}).

\subsection{Stage 3: Hybrid Fact Reconciliation and Domain-Specific Auto-Pairing}
\label{sec:stage3}

Stage 3 bridges neural semantic edits with the rigid structural definitions of the verification engine. To prevent neural hallucinations from eroding valid structural evidence, the reconciliation pipeline computes the finalized fact set $F_{\mathrm{final}}$ using two protective rules:
\begin{equation}
\label{eq:reconciliation_logic}
\begin{split}
F_{\mathrm{final}} &= \left( F_{\mathrm{AST}} \setminus \left\{ f \in \Delta^{-} \,\big|\, \neg\,\mathrm{ForceKeep}(f) \right\} \right) \\
&\quad \cup \left\{ f \in \Delta^{+} \,\big|\, \mathrm{SourceBacked}(f) \right\}
\end{split}
\end{equation}

\begin{itemize}[leftmargin=1em]
    \item \textbf{The Source-Backed Filter:} Evaluates to true only if a neural-proposed addition ($\Delta^+$) maps to a valid line number within the code and strictly matches the literal text, ensuring the LLM does not invent missing security checks.
    \item \textbf{The Force-Keep Filter:} Acts as a defensive lock against inaccurate LLM deletions by leveraging the syntactic confidence score $\gamma \in \{\mathrm{High}, \mathrm{Low}\}$ assigned during Stage~1. If a candidate fact carries this high syntactic certainty, the filter strictly prohibits the LLM from arbitrarily removing it.
\end{itemize}

\begin{table}[htbp]
\centering
\caption{Domain-specific auto-pairing strategies used to assemble verifiable claims for the five supported CWE classes.}
\label{tab:autopair}
\resizebox{\columnwidth}{!}{%
\begin{tabular}{@{}lp{0.9\linewidth}@{}}
\toprule
\textbf{CWE} & \textbf{Auto-Pairing Strategy} \\
\midrule
476 & For each dereference, compute the protection scope of each same-named nullity guard and pair when the guard precedes and structurally encloses the dereference (\texttt{earlyReturn}, \texttt{gotoError}, and \texttt{assertion} guard subsequent flow, whereas \texttt{explicitCheck} guards only its consequent block). Unguarded return-value dereferences are forced into mandatory claims, and a regex fallback recovers obfuscated guards. \\
\midrule
120 & For each write, search for a preceding boundary check sharing the same (or fuzzy-matched) buffer name and logically corresponding size/index variables. If no line-constrained match exists, fall back to a permissive name-only matching heuristic. \\
\midrule
125 & For each read, pair with a chronologically preceding check, actively proving the relation via shared buffer names, capacity expressions (\texttt{end - p >= N}, \texttt{len - off >= N}), or short-circuit bounds (\texttt{len < N || memcmp(...)}). Localized constant and symbolic index proofs are also supported. \\
\midrule
416 & Each use is natively anchored to its preceding release; scan the intervening span for a valid lifetime repair (nullification, reassignment, reference acquisition, ownership transfer, or escape guard). Pairings separated by a return, goto, or error-path barrier are aborted as unreachable unless the release is a specialized callback. \\
\midrule
415 & Group releases by normalized object identity and scan in source order; pair each release with the first path-feasible subsequent release on the same object. Intervening state resets, reallocations, or ownership repairs are appended as protections, while mutually exclusive branches and barrier-separated pairs are filtered out. \\
\bottomrule
\end{tabular}%
}
\end{table}

Applying the reconciliation logic from \autoref{eq:reconciliation_logic} to our running example, the LLM explicitly requested the removal of the post-release use at line 6 (\autoref{lst:stage2_interaction}). However, this deletion is intercepted by the $\mathrm{ForceKeep}$ filter. Because the invalidation at line 3 and the use at line 6 correspond to high-confidence AST nodes without any intervening lifetime repair ($P_o = \emptyset$), the critical obligations are retained.

Following reconciliation, the system transitions from disconnected facts to structured evidence by executing a domain-specific \textbf{auto-pairing} process. This step links retained dangerous operations with available protective facts to construct cohesive, verifiable claims. The pairing logic is strictly tailored to the semantics of each weakness class and summarized in \autoref{tab:autopair}.
For our running example, the retained invalidation at line 3 and use at line 6 are auto-paired by the CWE-416 logic into a verifiable claim $c = \langle\, o{:}(\text{Release: Line 3},\ \text{Use: Line 6}),\ P_o{=}\emptyset \,\rangle$, whose empty protection set leaves the underlying obligation undischarged.

\begin{lstlisting}[
    language=caml,
    caption={Abridged Lean 4 Verification Model mapping to $\mathcal{M} = \langle E, P, O \rangle$.}, 
    label={lst:stage4_lean},
    basicstyle=\ttfamily\footnotesize,
    frame=single,
    keywordstyle=\bfseries
]
-- Entity Definition (E)
def ptr_sk : Ptr := { name := "sk" }

-- Obligation/Claim Definition (O) retained after Stage 3
def claim_0 : LifetimeClaim := {
  use     := { ptr := ptr_sk, location := "line 6: release_sock(sk)" },
  release := { ptr := ptr_sk, location := "line 3: sco_send_frame(...)" },
  checks  := [] -- Protections (P_o): Empty, no valid repair dominates the use
}

-- The formal model fails to discharge claim_0, yielding VULNERABLE.
\end{lstlisting}

\subsection{Stage 4: Kernel-Checked Symbolic Verification in Lean 4}
\label{sec:stage4}

The auto-paired claims and tracked entities are automatically compiled into a declarative formal verification script and discharged by a localized Lean 4 kernel instance. Crucially, Lean does not re-parse the raw C/C++ source code; instead, it mathematically evaluates the structural consistency of the abstracted Verification Model $\mathcal{M}$. The script materializes one explicit definition per entity and per protection fact, and encodes each obligation as a distinct claim definition.

For every claim, the verifier evaluates the discharge relation of \autoref{eq:discharge_formal} under total, kernel-checked proofs over the encoded security contract. Each claim therefore resolves to a per-claim outcome in $\{\mathtt{proved}, \mathtt{unproved}, \mathtt{error}\}$: it is $\mathtt{proved}$ when a type-compatible protection bound to the same entity structurally covers the obligation, $\mathtt{unproved}$ when no admissible protection in its set covers the obligation, and $\mathtt{error}$ when it references an undeclared entity. The latter does not arise from well-formed input; it signals an encoding or environment fault rather than a property of the code, and lies outside the normal verdict path. The model emits the symbolic verdict $V_{Lean} = \mathtt{VULNERABLE}$ whenever at least one claim remains $\mathtt{unproved}$, and $\mathtt{SAFE}$ only when every claim is discharged. Importantly, it is the full per-claim outcome vector, not merely the aggregate verdict, that is forwarded to Stage~6, since the downstream grader reasons specifically over the residual set of unproven claims.

For the \texttt{sco\_sock\_sendmsg} running example, the generated Lean script (illustrated in \autoref{lst:stage4_lean}) models the obligation on \texttt{sk}. Because the parsed model contains no valid lifetime repair ($P_o = \emptyset$ for \texttt{sk}), the claim cannot be discharged, is marked \texttt{unproved}, and yields a symbolic $\mathtt{VULNERABLE}$ verdict. Conversely, for the patched code (\autoref{lst:patched}), decoupling the user copy from the locked region removes the obligation on \texttt{sk} entirely; with no obligation left to discharge, the model for \texttt{sk} is empty and the verifier reports $\mathtt{SAFE}$.

\subsection{Stage 5: Neuro-Symbolic Refinement}
\label{sec:stage5}

To mitigate the rigid failure modes of one-pass static verification, \tool{} incorporates an automated neuro-symbolic refinement step. It is invoked at most once per sample rather than iterated to convergence, and is conditionally executed under two programmatic triggers:

\begin{enumerate}
    \item \textbf{Conflict-Driven Refinement:} If the structural verdict from the Lean kernel ($V_{Lean}$) contradicts the semantic verdict returned by the neural network ($V_{LLM}$), the pipeline initiates a single refinement round. The prompt explicitly surfaces the specific unproved symbolic claims, forcing the LLM to re-examine the source text within this adversarial framing.
    \item \textbf{Empty-Model Retry Trigger:} If Stage 3 yields a completely vacuous obligation set ($|O| = 0$) but the source code contains strong lexical markers of dynamic resource lifecycles (e.g., \texttt{free}, \texttt{unref}, \texttt{destroy}), a retry is invoked to search specifically for hidden release paths. This trigger is specialized for the lifecycle-oriented weakness classes CWE-416 and CWE-415, where defects are frequently mediated by callees, refcounts, or callbacks that escape purely syntactic extraction.
\end{enumerate}

\subsection{Stage 6: Evidence Quality Grading and Adjudication}
\label{sec:stage6}

Recognizing that an \texttt{unproved} symbolic claim may occasionally stem from syntactic extraction noise, and that an LLM classification can suffer from generalized reasoning bias, the final component executes an evidence-aware fusion sequence rather than echoing $V_{Lean}$ directly.

\noindent\textbf{Symbolic grade.} The symbolic grade $Q_{Lean} \in \{\mathtt{strong}, \mathtt{weak}, \mathtt{noisy}, \mathtt{none}\}$ is derived exclusively from the set of \texttt{unproved} claims emitted by Stage~4, through two complementary pattern-matching passes. If Stage~4 leaves no unproved claim, the grade is immediately $\mathtt{none}$. Otherwise, each unproved claim is tested against two lexicons: (i) a \emph{noise} lexicon of structurally benign artifacts, such as fixed-size copies like \texttt{strcpy(a, b, 4)}, primitive scalar writes, or static compile-time lookup-table accesses; and (ii) a \emph{high-risk} lexicon, namely the same CWE-specific risk patterns that earn the syntactic confidence score $\gamma = \mathrm{High}$ during the Stage~1 tagging pass. The grade is then resolved by two group-level conditions: if \emph{every} unproved claim matches the noise lexicon, the grade is $\mathtt{noisy}$; if \emph{at least one} unproved claim matches the high-risk lexicon, the grade is $\mathtt{strong}$; otherwise the grade defaults to $\mathtt{weak}$. Intuitively, $\mathtt{noisy}$ indicates that the residual obligations are all explainable as extraction artifacts, $\mathtt{strong}$ indicates that at least one obligation is anchored to a $\gamma = \mathrm{High}$ sink, and $\mathtt{weak}$ covers the middle ground.

\noindent\textbf{Neural grade.} Concurrently, the system maps the textual reasoning block to a neural quality grade $Q_{LLM} \in \{\mathtt{strong}, \mathtt{weak}, \mathtt{none}\}$, assessing whether the model grounded its logic in literal variable identifiers and the correct CWE semantics, rather than cross-CWE hallucination.

\noindent\textbf{Adjudication.} The final binary categorization is produced by an asymmetric adjudication matrix (\autoref{tab:adjudication_expanded}), keyed on the tuple $(V_{Lean}, Q_{Lean}, V_{LLM}, Q_{LLM}, |O|)$. The matrix prioritizes structural evidence when it is validated by syntax ($Q_{Lean} = \mathtt{strong}$), while deferring to neural semantics when the symbolic model is vacuous or noisy.

For the running example, the final state evaluates as follows:
\begin{itemize}[leftmargin=2em]
    \item \textbf{Neural State:} $V_{LLM} = \mathtt{SAFE}$, but graded $Q_{LLM} = \mathtt{none}$ due to missing the cross-call invalidation risk.
    \item \textbf{Symbolic State:} $V_{Lean} = \mathtt{VULNERABLE}$, graded $Q_{Lean} = \mathtt{strong}$ because the single unproved claim matches a high-risk post-release pattern anchored to syntactically verified, distinct AST boundaries.
\end{itemize}
Resolving this configuration through the adjudication matrix (\autoref{tab:adjudication_expanded}, Row 1), the system correctly overrides the neural false negative. The pipeline outputs a final \textbf{\texttt{VULNERABLE}} verdict, successfully verifying the defect structurally and matching the ground truth.

\begin{table*}[t]
\centering
\caption{Evidence-aware adjudication matrix for CWE-416. While the overarching adjudication principles remain consistent, specific matrix configurations may vary slightly across other CWE classes to accommodate differences in AST extraction volumes and structural properties. Matrices for other CWEs are provided in \textit{our anonymous artifact}.}
\label{tab:adjudication_expanded}
\resizebox{\linewidth}{!}{
\begin{tabular}{@{}llll p{0.55\linewidth}@{}}
\toprule
\textbf{$V_{Lean}$ Verdict ($Q_{Lean}$)} &
\textbf{$V_{LLM}$ Verdict ($Q_{LLM}$)} &
\textbf{Obligation State ($|O|$)} &
\textbf{Final Output} &
\textbf{Pipeline Adjudication Rationale} \\
\midrule
$\mathtt{VULNERABLE}$ ($\mathtt{strong}$) &
$\mathtt{SAFE}$ ($\mathtt{none}$) &
$|O| \geq 1$ &
\textbf{\texttt{VULNERABLE}} &
Lean isolates a verifiable structural path that the neural model failed to dismiss with sufficient evidence. Overrides neural false negatives. \\
\midrule
$\mathtt{SAFE}$ ($\mathtt{none}$) &
$\mathtt{VULNERABLE}$ ($\mathtt{strong}$) &
$|O| = 0$ &
\textbf{\texttt{VULNERABLE}} &
The symbolic model is vacuous due to extraction limits; the system defers to high-confidence neural source evidence. \\
\midrule
$\mathtt{VULNERABLE}$ ($\mathtt{noisy}$) &
$\mathtt{SAFE}$ ($\mathtt{none}$) &
$|O| \geq 1$ &
\textbf{\texttt{SAFE}} &
The unproved claim is identified as a known syntactic artifact; the system defers to the neural safe verdict. \\
\midrule
$\mathtt{VULNERABLE}$ ($\mathtt{strong} \mid \mathtt{weak}$) &
$\mathtt{VULNERABLE}$ ($\mathtt{strong} \mid \mathtt{weak} \mid \mathtt{none}$) &
$|O| \geq 1$ &
\textbf{\texttt{VULNERABLE}} &
Non-noisy Lean evidence preserves the vulnerable consensus. \\
\midrule
$\mathtt{VULNERABLE}$ ($\mathtt{noisy}$) &
$\mathtt{VULNERABLE}$ ($\mathtt{none}$) &
$|O| \geq 1$ &
\textbf{\texttt{SAFE}} &
Mutual failure to produce high-confidence evidence triggers a safety downgrade to minimize false positives. \\
\midrule
$\mathtt{SAFE}$ ($\mathtt{none}$) &
$\mathtt{SAFE}$ ($\mathtt{none}$) &
$|O| \geq 0$ &
\textbf{\texttt{SAFE}} &
Consensus safety classification across the symbolic and neural blocks. \\
\bottomrule
\end{tabular}
}
\end{table*}

\section{Evaluation}
\label{sec:evaluation}

\subsection{Research Questions}
We aim to answer the following research questions (RQs):
\begin{itemize}[leftmargin=1em]
  \item \textbf{RQ1 (Effectiveness):} How effective is  \tool{} compared with pure LLM and agent baselines?
  \item \textbf{RQ2 (Ablation):} How much does each stage contribute?
  \item \textbf{RQ3 (Stability):} How reproducible are its results across independent runs?
  \item \textbf{RQ4 (Cost):} What are its token and time overheads?
\end{itemize}

\subsection{Experimental Setup}
\noindent\textbf{Datasets.} 
We first construct a large-scale, high-quality vulnerability dataset. We draw from three recent high-quality sources (PrimeVul~\cite{PrimeVul}, ReposVul~\cite{ReposVul}, and R2Vul~\cite{R2Vul}) that address known data quality issues in earlier datasets~\cite{BigVUl,CVEfixes,Chen2023DiverseVul}, such as label noise and tangled patches~\cite{li2025cleanvulautomaticfunctionlevelvulnerability}. We merge these three datasets and apply three filtering steps: (1) scope filtering, which excludes non-source files and test files; (2) length constraints, which discard samples exceeding a context window of 16,384 tokens to fit mainstream LLMs; and (3) de-duplication, which removes identical samples via MD5 hashing on normalized code~\cite{SECVULEVAL,SecLLMHolmes}. The resulting dataset comprises 5,078 CVEs across 149 CWE types.
From this dataset, we carefully select five representative CWE types for our experiments. We apply the following principles for selection: (1) non-logical vulnerabilities in C/C++ with clear semantics that are amenable to formal verification in Lean; (2) among the CWE Top 25 most dangerous software weaknesses~\cite{mitre_cwe_top25}; and (3) at least 100 samples per type for statistical reliability. The final experimental dataset comprises 138 samples for CWE-476, 116 for CWE-120, 183 for CWE-416, 255 for CWE-125, and 109 for CWE-415, totaling 801 samples (403 vulnerable and 398 non-vulnerable). Within each CWE type, we maintain an approximately balanced 1:1 ratio of vulnerable to non-vulnerable samples.
To enable agents to retrieve vulnerability-relevant interprocedural context locally, we clone the source code of 577 distinct repository versions via the GitHub commit links associated with each CVE.

\noindent\textbf{Baselines.}
All neural components, in both \tool{} and the baselines, use the same backend, DeepSeek-V4-Pro~\cite{deepseekai2026deepseekv4}, at temperature zero, which isolates the effect of the pipeline from differences in the underlying model. We compare against three baselines: a \emph{pure LLM} that reads the function and emits a verdict directly, and two agentic settings, Codex~\cite{openai2021codexweb} and Claude  Code~\cite{anthropic2026claude-code}, that inspect the repository context and reason over multiple steps. For each backend, ``\textit{\textbf{Ours}}'' denotes \tool{} operating under that setting. 

\noindent\textbf{Metrics.}
We report accuracy (ACC), F1, precision (Prec), and recall (Rec), computed with vulnerable code as the positive class. Because the practical objective of vulnerability detection is to surface genuinely vulnerable functions while keeping false alarms low, we center the analysis on \emph{F1}, which jointly summarizes the precision and recall of the positive class.

\subsection{RQ1: Effectiveness}
\label{sec:rq1}

\autoref{tab:rq1} compares \tool{} against the pure LLM baseline and the two agentic baselines across all five CWE classes and three backends. Across all fifteen dataset and backend combinations, \tool{} improves F1 over the corresponding baseline, without exception.

The dominant driver of the improvement is recall. The baselines, and the agentic baselines in particular, are \textbf{\textit{strongly precision-biased}}: they flag only the most blatant cases and miss the majority of true defects. This is most extreme on the lifecycle classes, for example, CWE-416 under Codex (Agent precision $0.9286$ but recall only $0.3171$) and CWE-415 under Claude Code (Agent precision $0.9167$ but recall only $0.2075$). By construction, \tool{} rebalances this regime: its deterministic auto-pairing and evidence-aware adjudication recover defects that the neural model dismisses, lifting recall substantially while incurring only a modest precision cost. On CWE-415, the F1 gain reaches $+0.20$ under Claude Code ($0.3385 \to 0.5432$) and $+0.19$ under the pure LLM ($0.4324 \to 0.6237$), with recall roughly doubling in both cases. On CWE-416, the gain is likewise recall-driven: under Codex, recall rises from $0.3171$ to $0.5062$ and F1 from $0.4727$ to $0.6165$ ($+0.14$), while under the pure LLM F1 climbs from $0.3652$ to $0.5000$. On CWE-476, the recall climbs above $0.84$ across all backends, raising F1 by $0.03$ to $0.06$. The pattern is consistent with the design goal stated in \autoref{sec:approach}: \tool{} converts the LLM from a final classifier into a constrained reviewer whose conservative deletions are intercepted by the structural pipeline. This rebalancing is a deliberate trade-off: converting borderline cases into flags can slightly lower positive-class precision (e.g., CWE-476 pure LLM, $0.5914\to0.5812$), which we accept because a missed vulnerability is costlier than an extra function queued for review. Notably, the benefit is not confined to the expensive agentic harnesses: the cheapest backend, the pure LLM, which is also the most precision-starved baseline, enjoys some of the largest improvements ($+0.19$ on CWE-415, $+0.13$ on CWE-416).

\begin{table}[htbp]
\centering
\caption{\textbf{RQ1: Effectiveness} of \tool{} against the pure LLM and agentic (Codex, Claude Code) baselines. We treat \textbf{F1 as the primary metric}: it balances precision and recall over the positive (vulnerable) class, which is the quantity of practical interest. For CWE-120, CWE-416, and CWE-476, we report the best of three runs by F1 (full statistics in \autoref{tab:rq3_meanstd}); CWE-125 and CWE-415 are single runs due to cost constraints (see \autoref{sec:rq3}).}
\label{tab:rq1}
\resizebox{0.9\linewidth}{!}{%
\begin{tabular}{@{}lll cccc@{}}
\toprule
\textbf{Dataset} & \textbf{Backend} & \textbf{Method} & \textbf{ACC} & \textbf{F1} & \textbf{Prec} & \textbf{Rec} \\
\midrule
\multirow{6}{*}{\shortstack[l]{\textbf{CWE-120}\\\scriptsize $N{=}116$}}
  & \multirow{2}{*}{Claude Code}   & Agent & 0.6696 & 0.6607 & 0.7400 & 0.5968 \\
  &                           & \ob\textbf{Ours} & \ob\textbf{0.7080} & \ob\fw{0.7080} & \ob\textbf{0.7547} & \ob\textbf{0.6667} \\
\cmidrule(l){2-7}
  & \multirow{2}{*}{Codex}    & Agent & 0.6293 & 0.6261 & 0.6792 & 0.5806 \\
  &                           & \ob\textbf{Ours} & \ob\textbf{0.6983} & \ob\fw{0.6957} & \ob\textbf{0.7547} & \ob\textbf{0.6452} \\
\cmidrule(l){2-7}
  & \multirow{2}{*}{Pure LLM} & LLM   & 0.6379 & 0.6379 & 0.6852 & 0.5968 \\
  &                           & \ob\textbf{Ours} & \ob\textbf{0.6638} & \ob\fw{0.6929} & \ob\textbf{0.6769} & \ob\textbf{0.7097} \\
\midrule
\multirow{6}{*}{\shortstack[l]{\textbf{CWE-416}\\\scriptsize $N{=}183$}}
  & \multirow{2}{*}{Claude Code}   & Agent & 0.7198 & 0.6047 & 0.8298 & 0.4756 \\
  &                           & \ob\textbf{Ours} & \ob\textbf{0.7432} & \ob\fw{0.6667} & \ob\textbf{0.7966} & \ob\textbf{0.5732} \\
\cmidrule(l){2-7}
  & \multirow{2}{*}{Codex}    & Agent & 0.6831 & 0.4727 & 0.9286 & 0.3171 \\
  &                           & \ob\textbf{Ours} & \ob\textbf{0.7198} & \ob\fw{0.6165} & \ob\textbf{0.7885} & \ob\textbf{0.5062} \\
\cmidrule(l){2-7}
  & \multirow{2}{*}{Pure LLM} & LLM   & 0.6011 & 0.3652 & 0.6364 & 0.2561 \\
  &                           & \ob\textbf{Ours} & \ob\textbf{0.6503} & \ob\fw{0.5000} & \ob\textbf{0.6957} & \ob\textbf{0.3902} \\
\midrule
\multirow{6}{*}{\shortstack[l]{\textbf{CWE-476}\\\scriptsize $N{=}138$}}
  & \multirow{2}{*}{Claude Code}   & Agent & 0.6107 & 0.6792 & 0.6279 & 0.7397 \\
  &                           & \ob\textbf{Ours} & \ob\textbf{0.6667} & \ob\fw{0.7356} & \ob\textbf{0.6531} & \ob\textbf{0.8421} \\
\cmidrule(l){2-7}
  & \multirow{2}{*}{Codex}    & Agent & 0.6397 & 0.6918 & 0.6471 & 0.7432 \\
  &                           & \ob\textbf{Ours} & \ob\textbf{0.6522} & \ob\fw{0.7273} & \ob\textbf{0.6400} & \ob\textbf{0.8421} \\
\cmidrule(l){2-7}
  & \multirow{2}{*}{Pure LLM} & LLM   & 0.5725 & 0.6509 & 0.5914 & 0.7237 \\
  &                           & \ob\textbf{Ours} & \ob\textbf{0.5870} & \ob\fw{0.7047} & \ob\textbf{0.5812} & \ob\textbf{0.8947} \\
\midrule
\multirow{6}{*}{\shortstack[l]{\textbf{CWE-125}\\\scriptsize $N{=}255$}}
  & \multirow{2}{*}{Claude Code}   & Agent & 0.6510 & 0.6510 & 0.6640 & 0.6385 \\
  &                           & \ob\textbf{Ours} & \ob\textbf{0.6510} & \ob\fw{0.7138} & \ob\textbf{0.6133} & \ob\textbf{0.8538} \\
\cmidrule(l){2-7}
  & \multirow{2}{*}{Codex}    & Agent & 0.6549 & 0.6641 & 0.6591 & 0.6692 \\
  &                           & \ob\textbf{Ours} & \ob\textbf{0.6392} & \ob\fw{0.7070} & \ob\textbf{0.6033} & \ob\textbf{0.8538} \\
\cmidrule(l){2-7}
  & \multirow{2}{*}{Pure LLM} & LLM   & 0.6314 & 0.6714 & 0.6154 & 0.7385 \\
  &                           & \ob\textbf{Ours} & \ob\textbf{0.6392} & \ob\fw{0.6892} & \ob\textbf{0.6145} & \ob\textbf{0.7846} \\
\midrule
\multirow{6}{*}{\shortstack[l]{\textbf{CWE-415}\\\scriptsize $N{=}109$}}
  & \multirow{2}{*}{Claude Code}   & Agent & 0.6055 & 0.3385 & 0.9167 & 0.2075 \\
  &                           & \ob\textbf{Ours} & \ob\textbf{0.6606} & \ob\fw{0.5432} & \ob\textbf{0.7857} & \ob\textbf{0.4151} \\
\cmidrule(l){2-7}
  & \multirow{2}{*}{Codex}    & Agent & 0.6055 & 0.3582 & 0.8571 & 0.2264 \\
  &                           & \ob\textbf{Ours} & \ob\textbf{0.6422} & \ob\fw{0.5063} & \ob\textbf{0.7692} & \ob\textbf{0.3774} \\
\cmidrule(l){2-7}
  & \multirow{2}{*}{Pure LLM} & LLM   & 0.6147 & 0.4324 & 0.7619 & 0.3019 \\
  &                           & \ob\textbf{Ours} & \ob\textbf{0.6789} & \ob\fw{0.6237} & \ob\textbf{0.7250} & \ob\textbf{0.5472} \\
\bottomrule
\end{tabular}}
\end{table}

\subsection{RQ2: Ablation}
\label{sec:rq2}

To isolate the contribution of each component, we ablate the two neuro-symbolic stages on CWE-120, CWE-416, and CWE-476 (\autoref{tab:rq2}). We read precision as a proxy for false-alarm control, and recall for missed defects.

\noindent\textbf{(1) The deterministic translator is indispensable for verifiability.} Under w/o AST the success rate stays close to $100\%$ across all backends, because the deterministic translator always emits well-formed Lean. Under w/o Translation, asking the LLM to write Lean directly degrades the success rate sharply: Claude Code fails most often on CWE-120 ($82/116$, $29\%$ rejected), while the pure LLM backend (DeepSeek-V4-Pro) is the most reliable ($110/116$, $178/183$, $133/138$). On these three classes, generating valid kernel-checked artifacts is therefore not something the LLM can be trusted to do unaided.

\noindent\textbf{(2) Translation is a key driver of recall.} Removing the translator collapses recall on the agentic backends: on CWE-416 it drops from $0.5732$ to $0.2754$ under Claude Code and from $0.5062$ to $0.2368$ under Codex, and on CWE-120 under Claude Code it falls from $0.6667$ to $0.4500$. On these backends, the directly generated Lean is heavily precision-biased, marking only the most obvious unsafe cases (precision $0.7917$ and $0.7500$ on CWE-416) and missing the complex configurations that require precise guard and claim pairing. The pure LLM backend does not show this clean precision bias (for instance, on CWE-416, both precision and recall are low at $0.4412$ and $0.3750$), indicating that the failure mode of direct generation is backend-dependent. The deterministic \texttt{auto\_pair} logic inside the translator is what recovers the missed cases on the agentic backends.

\noindent\textbf{(3) The AST stage raises precision and overall F1.} Across all three classes, the structural extractor matches or exceeds w/o AST on F1 for every backend, including several near-ties (e.g., on CWE-120, $0.6957$ vs.\ $0.6942$ under Codex; on CWE-476/pure LLM, an essentially flat $0.7047$ vs.\ $0.7053$). Its consistent role is to curb the indiscriminate ``flag-everything'' regime of direct extraction. This is clearest on CWE-476, where w/o AST marks nearly everything vulnerable (recall $1.0000$/$0.9867$/$0.9605$ under Codex/Claude Code/pure LLM) at low precision; the AST stage trims these \emph{false} positives, lifting precision (e.g., $0.5522\to0.6531$ under Claude Code). On the lifecycle class CWE-416 the precision gain is largest ($0.5816\to0.7966$ under Claude Code, $0.6216\to0.7885$ under Codex); here it also trims some \emph{true} positives, lowering recall ($0.6951\to0.5732$ under Claude Code), yet net F1 still improves for every backend.

\begin{table}[t]
\centering
\caption{\textbf{RQ2: Ablation} of the two neuro-symbolic stages (single run). \textbf{Ours} is the complete pipeline. \textbf{w/o AST} removes the Stage~1 structural extractor (the LLM identifies guards and dereferences from source). \textbf{w/o Translation} removes the deterministic Stage~4 translator (the LLM writes Lean directly). \textbf{Success} is the number of samples whose Lean script compiled and verified out of the total.}
\label{tab:rq2}
\resizebox{0.9\linewidth}{!}{%
\begin{tabular}{@{}lll c cccc@{}}
\toprule

\textbf{Dataset} & \textbf{Backend} & \textbf{Method} & \textbf{Success} & \textbf{ACC} & \textbf{F1} & \textbf{Prec} & \textbf{Rec} \\
\midrule
\multirow{9}{*}{\shortstack[l]{\textbf{CWE-120}\\\scriptsize $N{=}116$}}
  & \multirow{3}{*}{Claude Code}   & \ob\textbf{Ours}        & \ob116/116 & \ob\textbf{0.7080} & \ob\textbf{0.7080} & \ob\textbf{0.7547} & \ob\textbf{0.6667} \\
  &                           & w/o AST              & 115/116 & 0.6261 & 0.6195 & 0.6731 & 0.5738 \\
  &                           & w/o Translation      & \fail{82/116} & 0.5732 & 0.5070 & 0.5806 & 0.4500 \\
\cmidrule(l){2-8}
  & \multirow{3}{*}{Codex}    & \ob\textbf{Ours}        & \ob116/116 & \ob\textbf{0.6983} & \ob\textbf{0.6957} & \ob\textbf{0.7547} & \ob\textbf{0.6452} \\
  &                           & w/o AST              & 116/116 & 0.6810 & 0.6942 & 0.7119 & 0.6774 \\
  &                           & w/o Translation      & \fail{100/116} & 0.6600 & 0.6222 & 0.7368 & 0.5385 \\
\cmidrule(l){2-8}
  & \multirow{3}{*}{Pure LLM} & \ob\textbf{Ours}        & \ob116/116 & \ob\textbf{0.6638} & \ob\textbf{0.6929} & \ob\textbf{0.6769} & \ob\textbf{0.7097} \\
  &                           & w/o AST              & 116/116 & 0.6034 & 0.6230 & 0.6333 & 0.6129 \\
  &                           & w/o Translation      & \fail{110/116} & 0.6455 & 0.6929 & 0.6471 & 0.7458 \\
\midrule
\multirow{9}{*}{\shortstack[l]{\textbf{CWE-416}\\\scriptsize $N{=}183$}}
  & \multirow{3}{*}{Claude Code}   & \ob\textbf{Ours}        & \ob183/183 & \ob\textbf{0.7432} & \ob\textbf{0.6667} & \ob\textbf{0.7966} & \ob\textbf{0.5732} \\
  &                           & w/o AST              & 183/183 & 0.6393 & 0.6333 & 0.5816 & 0.6951 \\
  &                           & w/o Translation      & \fail{158/183} & 0.6519 & 0.4086 & 0.7917 & 0.2754 \\
\cmidrule(l){2-8}
  & \multirow{3}{*}{Codex}    & \ob\textbf{Ours}        & \ob183/183 & \ob\textbf{0.7198} & \ob\textbf{0.6165} & \ob\textbf{0.7885} & \ob\textbf{0.5062} \\
  &                           & w/o AST              & 183/183 & 0.6503 & 0.5897 & 0.6216 & 0.5610 \\
  &                           & w/o Translation      & \fail{171/183} & 0.6257 & 0.3600 & 0.7500 & 0.2368 \\
\cmidrule(l){2-8}
  & \multirow{3}{*}{Pure LLM} & \ob\textbf{Ours}        & \ob183/183 & \ob\textbf{0.6503} & \ob\textbf{0.5000} & \ob\textbf{0.6957} & \ob\textbf{0.3902} \\
  &                           & w/o AST              & 182/183 & 0.6154 & 0.4444 & 0.6364 & 0.3415 \\
  &                           & w/o Translation      & \fail{178/183} & 0.5056 & 0.4054 & 0.4412 & 0.3750 \\
\midrule
\multirow{9}{*}{\shortstack[l]{\textbf{CWE-476}\\\scriptsize $N{=}138$}}
  & \multirow{3}{*}{Claude Code}   & \ob\textbf{Ours}        & \ob138/138 & \ob\textbf{0.6667} & \ob\textbf{0.7356} & \ob\textbf{0.6531} & \ob\textbf{0.8421} \\
  &                           & w/o AST              & 137/138 & 0.5547 & 0.7081 & 0.5522 & 0.9867 \\
  &                           & w/o Translation      & \fail{118/138} & 0.5593 & 0.6438 & 0.5802 & 0.7231 \\
\cmidrule(l){2-8}
  & \multirow{3}{*}{Codex}    & \ob\textbf{Ours}        & \ob138/138 & \ob\textbf{0.6522} & \ob\textbf{0.7273} & \ob\textbf{0.6400} & \ob\textbf{0.8421} \\
  &                           & w/o AST              & 138/138 & 0.5725 & 0.7204 & 0.5630 & 1.0000 \\
  &                           & w/o Translation      & \fail{122/138} & 0.5820 & 0.6531 & 0.6316 & 0.6761 \\
\cmidrule(l){2-8}
  & \multirow{3}{*}{Pure LLM} & \ob\textbf{Ours}        & \ob138/138 & \ob\textbf{0.5870} & \ob\textbf{0.7047} & \ob\textbf{0.5812} & \ob\textbf{0.8947} \\
  &                           & w/o AST              & 138/138 & 0.5580 & 0.7053 & 0.5573 & 0.9605 \\
  &                           & w/o Translation      & \fail{133/138} & 0.5338 & 0.6771 & 0.5462 & 0.8904 \\
\bottomrule
\end{tabular}}
\end{table}

\subsection{RQ3: Stability}
\label{sec:rq3}

Repeating every condition is costly under the agentic backends (\autoref{sec:rq4}), so we concentrate the stability budget on CWE-120, CWE-416, and CWE-476, each run three times; CWE-125 and CWE-415 are therefore single runs. \autoref{tab:rq3_meanstd} reports the mean and standard deviation of each metric, and \autoref{tab:rq3_ttest} reports paired t-tests on the per-round F1 differences.

\noindent\textbf{Run-to-run variance is low.} Standard deviations stay small throughout (F1 below $0.04$), so a single run is representative. The variance of \tool{} is comparable to the baselines, for example, F1 std of $0.008$ versus $0.029$ on CWE-120 under Claude Code, confirming that the added pipeline introduces no extra instability. As the only stochastic element is the class-independent neural component, we expect this to carry over to the single-run CWE-125 and CWE-415.

\noindent\textbf{The F1 gains are statistically meaningful.} All nine combinations show a positive mean $\Delta\mathrm{F1}$, and 6 of 9 reach $p<0.05$ under the paired t-test ($\mathrm{df}=2$). The effect is strongest and most stable on CWE-416 (for example, $+0.149 \pm 0.006$ under Codex, $t=41.66$, $p=0.0006$). The three non-significant cases (two on CWE-120 and one on CWE-476 under Claude Code) all stem from larger inter-round variability that widens the confidence interval; with only three runs, even genuine improvements may miss the threshold, so these should be read as a conservative lower bound. Aggregating across conditions removes the small-sample fragility: all nine $\Delta\mathrm{F1}$ are positive (sign test $p\approx0.002$), so the direction of improvement is robust even where individual cells are underpowered.

\begin{table}[htbp]
\centering
\caption{\textbf{RQ3.1: Reproducibility.} Mean $\pm$ standard deviation over three independent runs. A lower standard deviation indicates better reproducibility.}
\label{tab:rq3_meanstd}
\resizebox{\linewidth}{!}{%
\begin{tabular}{@{}lll cccc@{}}
\toprule

\textbf{Dataset} & \textbf{Backend} & \textbf{Method} & \textbf{ACC} & \textbf{F1} & \textbf{Precision} & \textbf{Recall} \\
\midrule
\multirow{6}{*}{\shortstack[l]{\textbf{CWE-120}\\\scriptsize $N{=}116$}}
  & \multirow{2}{*}{Claude Code}   & Agent & $0.6473 \pm 0.022$ & $0.6318 \pm 0.029$ & $0.7166 \pm 0.018$ & $0.5654 \pm 0.035$ \\
  &                           & \ob\textbf{Ours} & \ob$\mathbf{0.6969 \pm 0.009}$ & \ob$\mathbf{0.6978 \pm 0.008}$ & \ob$\mathbf{0.7501 \pm 0.007}$ & \ob$\mathbf{0.6523 \pm 0.010}$ \\
\cmidrule(l){2-7}
  & \multirow{2}{*}{Codex}    & Agent & $0.6149 \pm 0.027$ & $0.5977 \pm 0.039$ & $0.6746 \pm 0.023$ & $0.5376 \pm 0.050$ \\
  &                           & \ob\textbf{Ours} & \ob$\mathbf{0.6810 \pm 0.014}$ & \ob$\mathbf{0.6683 \pm 0.020}$ & \ob$\mathbf{0.7518 \pm 0.013}$ & \ob$\mathbf{0.6022 \pm 0.030}$ \\
\cmidrule(l){2-7}
  & \multirow{2}{*}{Pure LLM} & LLM   & $0.6236 \pm 0.013$ & $0.6223 \pm 0.019$ & $0.6709 \pm 0.013$ & $0.5806 \pm 0.028$ \\
  &                           & \ob\textbf{Ours} & \ob$\mathbf{0.6408 \pm 0.020}$ & \ob$\mathbf{0.6661 \pm 0.027}$ & \ob$\mathbf{0.6618 \pm 0.018}$ & \ob$\mathbf{0.6720 \pm 0.052}$ \\
\midrule
\multirow{6}{*}{\shortstack[l]{\textbf{CWE-416}\\\scriptsize $N{=}183$}}
  & \multirow{2}{*}{Claude Code}   & Agent & $0.7093 \pm 0.015$ & $0.5729 \pm 0.036$ & $0.8430 \pm 0.014$ & $0.4350 \pm 0.043$ \\
  &                           & \ob\textbf{Ours} & \ob$\mathbf{0.7257 \pm 0.017}$ & \ob$\mathbf{0.6423 \pm 0.031}$ & \ob$\mathbf{0.7761 \pm 0.019}$ & \ob$\mathbf{0.5488 \pm 0.042}$ \\
\cmidrule(l){2-7}
  & \multirow{2}{*}{Codex}    & Agent & $0.6715 \pm 0.010$ & $0.4644 \pm 0.007$ & $0.8599 \pm 0.061$ & $0.3184 \pm 0.002$ \\
  &                           & \ob\textbf{Ours} & \ob$\mathbf{0.7190 \pm 0.008}$ & \ob$\mathbf{0.6111 \pm 0.008}$ & \ob$\mathbf{0.8020 \pm 0.027}$ & \ob$\mathbf{0.4939 \pm 0.011}$ \\
\cmidrule(l){2-7}
  & \multirow{2}{*}{Pure LLM} & LLM   & $0.5883 \pm 0.018$ & $0.3228 \pm 0.039$ & $0.6140 \pm 0.055$ & $0.2195 \pm 0.032$ \\
  &                           & \ob\textbf{Ours} & \ob$\mathbf{0.6393 \pm 0.011}$ & \ob$\mathbf{0.4896 \pm 0.018}$ & \ob$\mathbf{0.6693 \pm 0.023}$ & \ob$\mathbf{0.3862 \pm 0.019}$ \\
\midrule
\multirow{6}{*}{\shortstack[l]{\textbf{CWE-476}\\\scriptsize $N{=}138$}}
  & \multirow{2}{*}{Claude Code}   & Agent & $0.5841 \pm 0.028$ & $0.6426 \pm 0.034$ & $0.6102 \pm 0.022$ & $0.6792 \pm 0.053$ \\
  &                           & \ob\textbf{Ours} & \ob$\mathbf{0.6205 \pm 0.045}$ & \ob$\mathbf{0.7027 \pm 0.035}$ & \ob$\mathbf{0.6192 \pm 0.034}$ & \ob$\mathbf{0.8125 \pm 0.035}$ \\
\cmidrule(l){2-7}
  & \multirow{2}{*}{Codex}    & Agent & $0.6083 \pm 0.036$ & $0.6667 \pm 0.030$ & $0.6264 \pm 0.024$ & $0.7127 \pm 0.037$ \\
  &                           & \ob\textbf{Ours} & \ob$\mathbf{0.6375 \pm 0.027}$ & \ob$\mathbf{0.7140 \pm 0.021}$ & \ob$\mathbf{0.6305 \pm 0.016}$ & \ob$\mathbf{0.8231 \pm 0.029}$ \\
\cmidrule(l){2-7}
  & \multirow{2}{*}{Pure LLM} & LLM   & $0.5242 \pm 0.044$ & $0.6115 \pm 0.035$ & $0.5557 \pm 0.033$ & $0.6798 \pm 0.038$ \\
  &                           & \ob\textbf{Ours} & \ob$\mathbf{0.5459 \pm 0.036}$ & \ob$\mathbf{0.6770 \pm 0.024}$ & \ob$\mathbf{0.5566 \pm 0.021}$ & \ob$\mathbf{0.8640 \pm 0.027}$ \\
\bottomrule
\end{tabular}}
\end{table}

\begin{table}[htbp]
\centering
\caption{\textbf{RQ3.2: Significance of the F1 improvement.} For each combination, the per-round F1 differences $\Delta\mathrm{F1}=\mathrm{F1}_{\text{Ours}}-\mathrm{F1}_{\text{Baseline}}$ over the three runs are submitted to a paired t-test ($\mathrm{df}=2$). A larger $t$ means the improvement is more consistent; $p<0.05$ indicates at least $95\%$ confidence that the difference is not random. All nine $\Delta\mathrm{F1}$ are positive (sign test $p\approx0.002$).}
\label{tab:rq3_ttest}
\resizebox{0.9\linewidth}{!}{%
\begin{tabular}{@{}ll ccc c@{}}
\toprule

\textbf{Dataset} & \textbf{Backend} & \textbf{$\Delta$F1 (mean $\pm\sigma$)} & \textbf{$t$} & \textbf{$p$} & \textbf{Sig.} \\
\midrule
\multirow{3}{*}{\shortstack[l]{\textbf{CWE-120}\\\scriptsize $N{=}116$}}
  & Claude Code   & $+0.067 \pm 0.038$ & $3.05$  & $0.0928$ & \textcolor{gray}{No} \\
  & Codex    & $+0.071 \pm 0.036$ & $3.44$  & $0.0751$ & \textcolor{gray}{No} \\
  & Pure LLM & $+0.044 \pm 0.010$ & \textcolor{wingreen}{$\mathbf{7.77}$}  & \textcolor{wingreen}{$\mathbf{0.0162}$} & \textcolor{wingreen}{\textbf{Yes}} \\
\midrule
\multirow{3}{*}{\shortstack[l]{\textbf{CWE-416}\\\scriptsize $N{=}183$}}
  & Claude Code   & $+0.069 \pm 0.019$ & \textcolor{wingreen}{$\mathbf{6.21}$}  & \textcolor{wingreen}{$\mathbf{0.0250}$} & \textcolor{wingreen}{\textbf{Yes}} \\
  & Codex    & $+0.149 \pm 0.006$ & \textcolor{wingreen}{$\mathbf{41.66}$} & \textcolor{wingreen}{$\mathbf{0.0006}$} & \textcolor{wingreen}{\textbf{Yes}} \\
  & Pure LLM & $+0.167 \pm 0.028$ & \textcolor{wingreen}{$\mathbf{10.38}$} & \textcolor{wingreen}{$\mathbf{0.0092}$} & \textcolor{wingreen}{\textbf{Yes}} \\
\midrule
\multirow{3}{*}{\shortstack[l]{\textbf{CWE-476}\\\scriptsize $N{=}138$}}
  & Claude Code   & $+0.064 \pm 0.039$ & $2.87$  & $0.1030$ & \textcolor{gray}{No} \\
  & Codex    & $+0.049 \pm 0.013$ & \textcolor{wingreen}{$\mathbf{6.27}$}  & \textcolor{wingreen}{$\mathbf{0.0245}$} & \textcolor{wingreen}{\textbf{Yes}} \\
  & Pure LLM & $+0.066 \pm 0.011$ & \textcolor{wingreen}{$\mathbf{10.51}$} & \textcolor{wingreen}{$\mathbf{0.0089}$} & \textcolor{wingreen}{\textbf{Yes}} \\
\bottomrule
\end{tabular}}
\end{table}

\begin{table}[htbp]
\centering
\caption{\textbf{RQ4: Cost} of \tool{} per backend. Time is the per-sample mean ($\pm$ standard deviation across three runs); token columns are per-run totals averaged over three runs. Cache tokens are not applicable to the pure LLM setting.}
\label{tab:rq4}
\resizebox{\linewidth}{!}{%
\begin{tabular}{@{}ll cccc@{}}
\toprule

\textbf{Dataset} & \textbf{Backend} & \textbf{Time (s/sample)} & \textbf{Input Tokens (M)} & \textbf{Output Tokens (M)} & \textbf{Cache Tokens (M)} \\
\midrule
\multirow{3}{*}{\shortstack[l]{\textbf{CWE-120}\\\scriptsize $N{=}116$}}
  & Claude Code   & $205 \pm 2$  & $4.83 \pm 0.10$  & $1.64 \pm 0.05$ & $22.68 \pm 0.43$ \\
  & Codex    & $153 \pm 3$  & $24.68 \pm 3.52$ & $1.32 \pm 0.07$ & $19.92 \pm 3.29$ \\
  & Pure LLM & $73 \pm 15$  & $0.43 \pm 0.00$  & $0.50 \pm 0.01$ & n/a \\
\midrule
\multirow{3}{*}{\shortstack[l]{\textbf{CWE-416}\\\scriptsize $N{=}183$}}
  & Claude Code   & $129 \pm 5$  & $7.87 \pm 0.01$  & $2.46 \pm 0.05$ & $62.15 \pm 2.27$ \\
  & Codex    & $131 \pm 6$  & $48.28 \pm 0.82$ & $1.85 \pm 0.07$ & $40.30 \pm 0.73$ \\
  & Pure LLM & $37 \pm 1$   & $0.50 \pm 0.00$  & $0.53 \pm 0.01$ & n/a \\
\midrule
\multirow{3}{*}{\shortstack[l]{\textbf{CWE-476}\\\scriptsize $N{=}138$}}
  & Claude Code   & $276 \pm 17$ & $5.79 \pm 0.13$  & $2.28 \pm 0.07$ & $47.17 \pm 3.43$ \\
  & Codex    & $261 \pm 35$ & $48.87 \pm 2.28$ & $2.12 \pm 0.04$ & $40.45 \pm 2.07$ \\
  & Pure LLM & $104 \pm 9$  & $0.33 \pm 0.00$  & $0.78 \pm 0.00$ & n/a \\
\bottomrule
\end{tabular}}
\end{table}

\subsection{RQ4: Cost}
\label{sec:rq4}

\autoref{tab:rq4} reports the resource consumption of \tool{} under each backend on the three classes, run three times. The pure LLM backend is by far the cheapest, at $37$ to $104$ seconds per sample and well under one million input and output tokens per run, since it reasons over the function in isolation. 
This agentic spend does not buy proportional accuracy: Codex uses $5$--$8.5\times$ the input tokens of Claude Code without leading on F1, and the cheapest pure LLM backend is where \tool{} gains most. The verification machinery adds a bounded overhead, the price for the recall and F1 gains, and the kernel-checked evidence established in RQ1-RQ3.

\section{Threats to Validity}
\label{sec:discussion}

\noindent\textbf{Construct validity.}
We adopt F1 over the vulnerable class as our primary metric, which inherits the limitations of the benchmark labels we did not independently re-audit. A second threat is translation soundness: a verdict from \tool{} is only as trustworthy as the Lean encoding it checks. We mitigate this with a deterministic translator and report verification success (\autoref{tab:rq2}), but a verified script attests only to the modeled property, not the absence of all vulnerabilities. 

\noindent\textbf{Internal validity.}
LLM nondeterminism is the main internal threat. We repeat CWE-120, CWE-416, and CWE-476 over three runs and report mean$\pm$std, where \tool{} shows lower variance than the baselines; CWE-125 and CWE-415 are single runs and thus indicative. Prompts, temperatures, and tool versions are fixed across conditions so that differences reflect the method, not configuration drift.

\noindent\textbf{External validity.}
The dominant threat is the scope of the approach itself. \tool{} is not a general detector that ingests arbitrary code and routes it to the right weakness: each CWE class is handled by a hand-built specification with its own extraction rules and property templates, and the target class must be fixed in advance. Because different weaknesses expose structurally different obligations, this per-class customization is intrinsic to the current design, not an implementation shortcut, and we make no claim of generalization to unseen classes or to vulnerabilities requiring whole-program reasoning. We therefore position \tool{} as an exploratory prototype testing the feasibility of separating fact proposal from obligation discharge, not a deployable tool. Our five classes span distinct bug families but are far from exhaustive; within each, we use every usable sample to avoid cherry-picking. Lifting the per-class specification requirement, including automatic routing and a shared extraction layer, is the central direction for future work.

\section{Conclusion}
\label{sec:conclusion}

We presented \tool{}, a neuro-symbolic framework that separates supplying evidence, via an LLM semantic filter, from discharging safety obligations, via a deterministic Lean~4 checker and evidence-aware adjudication. Across five CWE classes and three backends, \tool{} improves F1 over the baselines in all fifteen settings.

\bibliographystyle{IEEEtranS}
\bibliography{main}

@misc{mitre2024cwe,
    author = {{The MITRE Corporation}},
    title = {Common Weakness Enumeration (CWE)},
    year = {2024},
    howpublished = {\url{https://cwe.mitre.org/}},
    note = {Accessed: 2026-06-26, Community-developed taxonomy of software security weaknesses}
}

@misc{nvd_linux_cwe476,
  author       = {{National Vulnerability Database}},
  title        = {Vulnerability Search Results: {Linux Kernel} and {CWE-476} ({NULL} Pointer Dereference)},
  year         = {2024},
  howpublished = {\url{https://nvd.nist.gov/vuln/search/results?form_type=Advanced&cpe_vendor=cpe%3A2.3%3Ao%3Alinux&cpe_product=cpe%3A2.3%3Ao%3Alinux%3Alinux_kernel&cwe_id=CWE-476}},
  note         = {Accessed: 2026-06-30}
}

@misc{cvedetails_linux_2024,
  author       = {{CVE Details}},
  title        = {Linux Kernel Vulnerability Statistics for 2024},
  year         = {2024},
  howpublished = {\url{https://www.cvedetails.com/vulnerability-list/vendor_id-33/product_id-47/year-2024/Linux-Linux-Kernel.html}},
  note         = {Accessed: 2026-06-30}
}

@misc{github2026codeql,
    author = {{GitHub, Inc.}},
    title = {CodeQL: Semantic Code Analysis Engine for Security Vulnerability Detection},
    year = {2026},
    howpublished = {\url{https://codeql.github.com/}},
    note = {Accessed: 2026-06-26, Relational static analysis tool supporting C/CWE vulnerability checking}
}

@misc{marjamaki2026cppcheck,
    author = {Daniel Marjamäki and Cppcheck Developers},
    title = {Cppcheck: Static Analysis Tool for C/C++},
    year = {2026},
    howpublished = {\url{https://cppcheck.sourceforge.io/}},
    note = {Accessed: 2026-06-26, Open-source static analyzer detecting undefined behavior, memory flaws and CWE vulnerabilities}
}

@article{cobbe,
  author       = {Karl Cobbe and
                  Vineet Kosaraju and
                  Mohammad Bavarian and
                  Mark Chen and
                  Heewoo Jun and
                  Lukasz Kaiser and
                  Matthias Plappert and
                  Jerry Tworek and
                  Jacob Hilton and
                  Reiichiro Nakano and
                  Christopher Hesse and
                  John Schulman},
  title        = {Training Verifiers to Solve Math Word Problems},
  journal      = {CoRR},
  volume       = {abs/2110.14168},
  year         = {2021},
  url          = {https://arxiv.org/abs/2110.14168},
  eprinttype   = {arXiv},
  eprint       = {2110.14168},
  bibsource    = {dblp computer science bibliography, https://dblp.org}
}

@inproceedings{deepseekprover,
  author       = {Huajian Xin and
                  Z. Z. Ren and
                  Junxiao Song and
                  Zhihong Shao and
                  Wanjia Zhao and
                  Haocheng Wang and
                  Bo Liu and
                  Liyue Zhang and
                  Xuan Lu and
                  Qiushi Du and
                  Wenjun Gao and
                  Haowei Zhang and
                  Qihao Zhu and
                  Dejian Yang and
                  Zhibin Gou and
                  Z. F. Wu and
                  Fuli Luo and
                  Chong Ruan},
  title        = {DeepSeek-Prover-V1.5: Harnessing Proof Assistant Feedback for Reinforcement
                  Learning and Monte-Carlo Tree Search},
  booktitle    = {The Thirteenth International Conference on Learning Representations,
                  {ICLR} 2025, Singapore, April 24-28, 2025},
  publisher    = {OpenReview.net},
  year         = {2025},
  url          = {https://openreview.net/forum?id=I4YAIwrsXa},
  bibsource    = {dblp computer science bibliography, https://dblp.org}
}

@inproceedings{
apollo,
title={{APOLLO}: Automated {LLM} and Lean Collaboration for Advanced Formal Reasoning},
author={Azim Ospanov and Farzan Farnia and Roozbeh Yousefzadeh},
booktitle={The Thirty-ninth Annual Conference on Neural Information Processing Systems},
year={2026},
url={https://openreview.net/forum?id=fxDCgOruk0}
}

@article{leanmeta,
  author       = {Sebastian Ullrich and
                  Leonardo de Moura},
  title        = {Beyond Notations: Hygienic Macro Expansion for Theorem Proving Languages},
  journal      = {Log. Methods Comput. Sci.},
  volume       = {18},
  number       = {2},
  year         = {2022},
  url          = {https://doi.org/10.46298/lmcs-18(2:1)2022},
  doi          = {10.46298/LMCS-18(2:1)2022},
  bibsource    = {dblp computer science bibliography, https://dblp.org}
}

@article{compcert,
  author       = {Xavier Leroy},
  title        = {Formal verification of a realistic compiler},
  journal      = {Commun. {ACM}},
  volume       = {52},
  number       = {7},
  pages        = {107--115},
  year         = {2009},
  url          = {https://doi.org/10.1145/1538788.1538814},
  doi          = {10.1145/1538788.1538814},
  bibsource    = {dblp computer science bibliography, https://dblp.org}
}

@inproceedings{sel4,
  author       = {Gerwin Klein and
                  Kevin Elphinstone and
                  Gernot Heiser and
                  June Andronick and
                  David A. Cock and
                  Philip Derrin and
                  Dhammika Elkaduwe and
                  Kai Engelhardt and
                  Rafal Kolanski and
                  Michael Norrish and
                  Thomas Sewell and
                  Harvey Tuch and
                  Simon Winwood},
  editor       = {Jeanna Neefe Matthews and
                  Thomas E. Anderson},
  title        = {se{L}4: formal verification of an OS kernel.},
  booktitle    = {Proceedings of the 22nd {ACM} Symposium on Operating Systems Principles
                  2009, {SOSP} 2009, Big Sky, Montana, USA, October 11-14, 2009},
  pages        = {207--220},
  publisher    = {{ACM}},
  year         = {2009},
  url          = {https://doi.org/10.1145/1629575.1629596},
  doi          = {10.1145/1629575.1629596},
  bibsource    = {dblp computer science bibliography, https://dblp.org}
}

@inproceedings{factalign,
  author       = {Zongze Jiang and
                  Ming Wen and
                  Ge Wen and
                  Hai Jin},
  title        = {Fact-Aligned and Template-Constrained Static Analyzer Rule Enhancement
                  with LLMs},
  booktitle    = {40th {IEEE/ACM} International Conference on Automated Software Engineering,
                  {ASE} 2025, Seoul, Korea, Republic of, November 16-20, 2025},
  pages        = {1642--1654},
  publisher    = {{IEEE}},
  year         = {2025},
  url          = {https://doi.org/10.1109/ASE63991.2025.00138},
  doi          = {10.1109/ASE63991.2025.00138},
  bibsource    = {dblp computer science bibliography, https://dblp.org}
}

@article{qlpro,
  title        = {QLPro: Automated Code Vulnerability Discovery via {LLM} and Static
                  Code Analysis Integration},
  journal      = {CoRR},
  volume       = {abs/2506.23644},
  year         = {2025},
  note         = {Withdrawn.},
  url          = {https://doi.org/10.48550/arXiv.2506.23644},
  doi          = {10.48550/ARXIV.2506.23644},
  eprinttype   = {arXiv},
  eprint       = {2506.23644},
  bibsource    = {dblp computer science bibliography, https://dblp.org}
}

@inproceedings{iris,
  author       = {Ziyang Li and
                  Saikat Dutta and
                  Mayur Naik},
  title        = {{IRIS:} LLM-Assisted Static Analysis for Detecting Security Vulnerabilities},
  booktitle    = {The Thirteenth International Conference on Learning Representations,
                  {ICLR} 2025, Singapore, April 24-28, 2025},
  publisher    = {OpenReview.net},
  year         = {2025},
  url          = {https://openreview.net/forum?id=9LdJDU7E91},
  bibsource    = {dblp computer science bibliography, https://dblp.org}
}

@article{deepseekcoder,
  author       = {Daya Guo and
                  Qihao Zhu and
                  Dejian Yang and
                  Zhenda Xie and
                  Kai Dong and
                  Wentao Zhang and
                  Guanting Chen and
                  Xiao Bi and
                  Y. Wu and
                  Y. K. Li and
                  Fuli Luo and
                  Yingfei Xiong and
                  Wenfeng Liang},
  title        = {DeepSeek-Coder: When the Large Language Model Meets Programming -
                  The Rise of Code Intelligence},
  journal      = {CoRR},
  volume       = {abs/2401.14196},
  year         = {2024},
  url          = {https://doi.org/10.48550/arXiv.2401.14196},
  doi          = {10.48550/ARXIV.2401.14196},
  eprinttype   = {arXiv},
  eprint       = {2401.14196},
  bibsource    = {dblp computer science bibliography, https://dblp.org}
}

@inproceedings{codebert,
  author       = {Zhangyin Feng and
                  Daya Guo and
                  Duyu Tang and
                  Nan Duan and
                  Xiaocheng Feng and
                  Ming Gong and
                  Linjun Shou and
                  Bing Qin and
                  Ting Liu and
                  Daxin Jiang and
                  Ming Zhou},
  editor       = {Trevor Cohn and
                  Yulan He and
                  Yang Liu},
  title        = {CodeBERT: {A} Pre-Trained Model for Programming and Natural Languages},
  booktitle    = {Findings of the Association for Computational Linguistics: {EMNLP}
                  2020, Online Event, 16-20 November 2020},
  series       = {Findings of {ACL}},
  volume       = {{EMNLP} 2020},
  pages        = {1536--1547},
  publisher    = {Association for Computational Linguistics},
  year         = {2020},
  url          = {https://doi.org/10.18653/v1/2020.findings-emnlp.139},
  doi          = {10.18653/V1/2020.FINDINGS-EMNLP.139},
  bibsource    = {dblp computer science bibliography, https://dblp.org}
}

@inproceedings{staticfp2,
  author       = {Joonyoung Park and
                  Inho Lim and
                  Sukyoung Ryu},
  editor       = {Laura K. Dillon and
                  Willem Visser and
                  Laurie A. Williams},
  title        = {Battles with false positives in static analysis of JavaScript web
                  applications in the wild},
  booktitle    = {Proceedings of the 38th International Conference on Software Engineering,
                  {ICSE} 2016, Austin, TX, USA, May 14-22, 2016 - Companion Volume},
  pages        = {61--70},
  publisher    = {{ACM}},
  year         = {2016},
  url          = {https://doi.org/10.1145/2889160.2889227},
  doi          = {10.1145/2889160.2889227},
  bibsource    = {dblp computer science bibliography, https://dblp.org}
}

@inproceedings{staticfp1,
  author       = {Hong Jin Kang and
                  Khai Loong Aw and
                  David Lo},
  title        = {Detecting False Alarms from Automatic Static Analysis Tools: How Far
                  are We?},
  booktitle    = {44th {IEEE/ACM} 44th International Conference on Software Engineering,
                  {ICSE} 2022, Pittsburgh, PA, USA, May 25-27, 2022},
  pages        = {698--709},
  publisher    = {{ACM}},
  year         = {2022},
  url          = {https://doi.org/10.1145/3510003.3510214},
  doi          = {10.1145/3510003.3510214},
  bibsource    = {dblp computer science bibliography, https://dblp.org}
}

@inproceedings{SecLLMHolmes,
  author        = {Saad Ullah and
                   Mingji Han and
                   Saurabh Pujar and
                   Hammond Pearce and
                   Ayse K. Coskun and
                   Gianluca Stringhini},
  bibsource     = {dblp computer science bibliography, https://dblp.org},
  booktitle     = {{IEEE} Symposium on Security and Privacy, {SP} 2024, San Francisco,
                   CA, USA, May 19-23, 2024},
  doi           = {10.1109/SP54263.2024.00210},
  pages         = {862--880},
  publisher     = {{IEEE}},
  title         = {{L}{L}{M}s {C}annot {R}eliably {I}dentify and {R}eason {A}bout {S}ecurity {V}ulnerabilities ({Y}et?): {A} {C}omprehensive {E}valuation, {F}ramework, and {B}enchmarks},
  url           = {https://doi.org/10.1109/SP54263.2024.00210},
  year          = {2024}
}

@misc{SECVULEVAL,
  archiveprefix = {arXiv},
  author        = {Md Basim Uddin Ahmed and Nima Shiri Harzevili and Jiho Shin and Hung Viet Pham and Song Wang},
  eprint        = {2505.19828},
  primaryclass  = {cs.SE},
  title         = {{S}ec{V}ul{E}val: {B}enchmarking {L}{L}{M}s for {R}eal-World {C}/{C}++ {V}ulnerability {D}etection},
  url           = {https://arxiv.org/abs/2505.19828},
  year          = {2025}
}

@article{Chen2023DiverseVul,
  author        = {Yizheng Chen and
                   Zhoujie Ding and
                   Xinyun Chen and
                   David A. Wagner},
  bibsource     = {dblp computer science bibliography, https://dblp.org},
  doi           = {10.48550/ARXIV.2304.00409},
  eprint        = {2304.00409},
  eprinttype    = {arXiv},
  journal       = {CoRR},
  title         = {{D}iverse{V}ul: {A} {N}ew {V}ulnerable {S}ource {C}ode {D}ataset for {D}eep {L}earning {B}ased {V}ulnerability {D}etection},
  url           = {https://doi.org/10.48550/arXiv.2304.00409},
  volume        = {abs/2304.00409},
  year          = {2023}
}

@article{CVEfixes,
  author        = {Guru Prasad Bhandari and
                   Amara Naseer and
                   Leon Moonen},
  bibsource     = {dblp computer science bibliography, https://dblp.org},
  eprint        = {2107.08760},
  eprinttype    = {arXiv},
  journal       = {CoRR},
  title         = {{C}{V}{E}fixes: {A}utomated {C}ollection of {V}ulnerabilities and {T}heir {F}ixes from {O}pen-Source {S}oftware},
  url           = {https://arxiv.org/abs/2107.08760},
  volume        = {abs/2107.08760},
  year          = {2021}
}

@inproceedings{BigVUl,
  author        = {Jiahao Fan and
                   Yi Li and
                   Shaohua Wang and
                   Tien N. Nguyen},
  bibsource     = {dblp computer science bibliography, https://dblp.org},
  booktitle     = {{MSR} '20: 17th International Conference on Mining Software Repositories,
                   Seoul, Republic of Korea, 29-30 June, 2020},
  doi           = {10.1145/3379597.3387501},
  editor        = {Sunghun Kim and
                   Georgios Gousios and
                   Sarah Nadi and
                   Joseph Hejderup},
  pages         = {508--512},
  publisher     = {{ACM}},
  title         = {{A} {C/C++} {C}ode {V}ulnerability {D}ataset with {C}ode {C}hanges and {CVE} {S}ummaries},
  url           = {https://doi.org/10.1145/3379597.3387501},
  year          = {2020}
}

@article{li2025cleanvulautomaticfunctionlevelvulnerability,
  author        = {Yikun Li and
                   Ting Zhang and
                   Ratnadira Widyasari and
                   Yan Naing Tun and
                   Huu Hung Nguyen and
                   Tan Bui and
                   Ivana Clairine Irsan and
                   Yiran Cheng and
                   Xiang Lan and
                   Han Wei Ang and
                   Frank Liauw and
                   Martin Weyssow and
                   Hong Jin Kang and
                   Eng Lieh Ouh and
                   Lwin Khin Shar and
                   David Lo},
  bibsource     = {dblp computer science bibliography, https://dblp.org},
  doi           = {10.48550/ARXIV.2411.17274},
  eprint        = {2411.17274},
  eprinttype    = {arXiv},
  journal       = {CoRR},
  title         = {{C}lean{V}ul: {A}utomatic {F}unction-Level {V}ulnerability {D}etection in {C}ode {C}ommits {U}sing {LLM} {H}euristics},
  url           = {https://doi.org/10.48550/arXiv.2411.17274},
  volume        = {abs/2411.17274},
  year          = {2024}
}

@article{R2Vul,
  author        = {Martin Weyssow and
                   Chengran Yang and
                   Junkai Chen and
                   Yikun Li and
                   Huihui Huang and
                   Ratnadira Widyasari and
                   Han Wei Ang and
                   Frank Liauw and
                   Eng Lieh Ouh and
                   Lwin Khin Shar and
                   David Lo},
  bibsource     = {dblp computer science bibliography, https://dblp.org},
  doi           = {10.48550/ARXIV.2504.04699},
  eprint        = {2504.04699},
  eprinttype    = {arXiv},
  journal       = {CoRR},
  title         = {{R}2{V}ul: {L}earning to {R}eason about {S}oftware {V}ulnerabilities with {R}einforcement {L}earning and {S}tructured {R}easoning {D}istillation},
  url           = {https://doi.org/10.48550/arXiv.2504.04699},
  volume        = {abs/2504.04699},
  year          = {2025}
}

@inproceedings{ReposVul,
  author        = {Xinchen Wang and
                   Ruida Hu and
                   Cuiyun Gao and
                   Xin{-}Cheng Wen and
                   Yujia Chen and
                   Qing Liao},
  bibsource     = {dblp computer science bibliography, https://dblp.org},
  booktitle     = {Proceedings of the 2024 {IEEE/ACM} 46th International Conference on
                   Software Engineering: Companion Proceedings, {ICSE} Companion 2024,
                   Lisbon, Portugal, April 14-20, 2024},
  doi           = {10.1145/3639478.3647634},
  pages         = {472--483},
  publisher     = {{ACM}},
  title         = {{R}epos{V}ul: {A} {R}epository-Level {H}igh-Quality {V}ulnerability {D}ataset},
  url           = {https://doi.org/10.1145/3639478.3647634},
  year          = {2024}
}

@article{PrimeVul,
  author        = {Yangruibo Ding and
                   Yanjun Fu and
                   Omniyyah Ibrahim and
                   Chawin Sitawarin and
                   Xinyun Chen and
                   Basel Alomair and
                   David A. Wagner and
                   Baishakhi Ray and
                   Yizheng Chen},
  bibsource     = {dblp computer science bibliography, https://dblp.org},
  doi           = {10.48550/ARXIV.2403.18624},
  eprint        = {2403.18624},
  eprinttype    = {arXiv},
  journal       = {CoRR},
  title         = {{V}ulnerability {D}etection with {C}ode {L}anguage {M}odels: {H}ow {F}ar {A}re {W}e?},
  url           = {https://doi.org/10.48550/arXiv.2403.18624},
  volume        = {abs/2403.18624},
  year          = {2024}
}

@misc{mitre_cve_metrics,
  author    = {{The MITRE Corporation}},
  title     = {CVE Metrics: Published CVE Records Statistics},
  year      = {2026},
  howpublished = {\url{https://www.cve.org/about/Metrics}}
}

@misc{mitre_cwe_top25,
  author    = {{The MITRE Corporation}},
  title     = {2025 CWE Top 25 Most Dangerous Software Weaknesses},
  year      = {2025},
  howpublished = {\url{https://cwe.mitre.org/top25/archive/2025/2025_cwe_top25}}
}

@misc{infer,
  title        = {Infer: A static analyzer for {Java}, {C}, {C++}, and {Objective-C}},
  author       = {{Meta}},
  howpublished = {\url{https://fbinfer.com/}},
  year         = {2026}
}

@misc{coverity,
  title        = {Coverity Static Analysis},
  author       = {{Synopsys}},
  howpublished = {\url{https://www.synopsys.com/software-integrity/security-testing/static-analysis-sast.html}},
  year         = {2026}
}

@misc{anthropic_mythos,
  title        = {Claude {Mythos}},
  author       = {{Anthropic}},
  howpublished = {\url{https://www.anthropic.com/claude/mythos}},
  year         = {2026}
}

@misc{google_bigsleep,
  title        = {From {Naptime} to {Big Sleep}: Using Large Language Models To Catch Vulnerabilities In Real-World Code},
  author       = {{Google Project Zero}},
  howpublished = {\url{https://projectzero.google/2024/10/from-naptime-to-big-sleep.html}},
  year         = {2024}
}

@misc{sakana_fugu,
  title        = {Sakana {Fugu}: One Model to Command Them All},
  author       = {{Sakana AI}},
  howpublished = {\url{https://sakana.ai/fugu-release/}},
  year         = {2026}
}

@article{10.1145/3695988,
author = {Hou, Xinyi and Zhao, Yanjie and Liu, Yue and Yang, Zhou and Wang, Kailong and Li, Li and Luo, Xiapu and Lo, David and Grundy, John and Wang, Haoyu},
title = {Large Language Models for Software Engineering: A Systematic Literature Review},
year = {2024},
issue_date = {November 2024},
publisher = {Association for Computing Machinery},
address = {New York, NY, USA},
volume = {33},
number = {8},
issn = {1049-331X},
url = {https://doi.org/10.1145/3695988},
doi = {10.1145/3695988},
journal = {ACM Trans. Softw. Eng. Methodol.},
month = dec,
articleno = {220},
numpages = {79}
}

@article{DBLP:journals/cacm/BesseyBCCFHHKME10,
  author       = {Al Bessey and
                  Ken Block and
                  Benjamin Chelf and
                  Andy Chou and
                  Bryan Fulton and
                  Seth Hallem and
                  Charles{-}Henri Gros and
                  Asya Kamsky and
                  Scott McPeak and
                  Dawson R. Engler},
  title        = {A few billion lines of code later: using static analysis to find bugs
                  in the real world},
  journal      = {Commun. {ACM}},
  volume       = {53},
  number       = {2},
  pages        = {66--75},
  year         = {2010},
  url          = {https://doi.org/10.1145/1646353.1646374},
  doi          = {10.1145/1646353.1646374},
  bibsource    = {dblp computer science bibliography, https://dblp.org}
}

@inproceedings{DBLP:conf/iclr/DeletangRGGWCCH23,
  author       = {Gr{\'{e}}goire Del{\'{e}}tang and
                  Anian Ruoss and
                  Jordi Grau{-}Moya and
                  Tim Genewein and
                  Li Kevin Wenliang and
                  Elliot Catt and
                  Chris Cundy and
                  Marcus Hutter and
                  Shane Legg and
                  Joel Veness and
                  Pedro A. Ortega},
  title        = {Neural Networks and the Chomsky Hierarchy},
  booktitle    = {The Eleventh International Conference on Learning Representations,
                  {ICLR} 2023, Kigali, Rwanda, May 1-5, 2023},
  publisher    = {OpenReview.net},
  year         = {2023},
  url          = {https://openreview.net/forum?id=WbxHAzkeQcn},
  bibsource    = {dblp computer science bibliography, https://dblp.org}
}

@article{10.1145/3769676,
author = {Xu, Hanxiang and Wang, Shenao and Li, Ningke and Wang, Kailong and Zhao, Yanjie and Chen, Kai and Yu, Ting and Liu, Yang and Wang, Haoyu},
title = {Large Language Models for Cyber Security: A Systematic Literature Review},
year = {2025},
publisher = {Association for Computing Machinery},
address = {New York, NY, USA},
issn = {1049-331X},
url = {https://doi.org/10.1145/3769676},
doi = {10.1145/3769676},
note = {Just Accepted},
journal = {ACM Trans. Softw. Eng. Methodol.},
month = sep
}

@inproceedings{10.1007/978-3-030-79876-5_37,
author = {Moura, Leonardo de and Ullrich, Sebastian},
title = {The Lean 4 Theorem Prover and Programming Language},
year = {2021},
isbn = {978-3-030-79875-8},
publisher = {Springer-Verlag},
address = {Berlin, Heidelberg},
url = {https://doi.org/10.1007/978-3-030-79876-5_37},
doi = {10.1007/978-3-030-79876-5_37},
booktitle = {Automated Deduction – CADE 28: 28th International Conference on Automated Deduction, Virtual Event, July 12–15, 2021, Proceedings},
pages = {625–635},
numpages = {11}
}

@misc{deepseekai2026deepseekv4,
    title={DeepSeek-V4: Towards Highly Efficient Million-Token Context Intelligence},
    author={DeepSeek-AI},
    year={2026},
    howpublished={\url{https://huggingface.co/deepseek-ai/DeepSeek-V4-Pro}}
}

@misc{anthropic2026claude-code,
    title={Claude Code: Agentic Coding Assistant},
    author={Anthropic},
    year={2026},
    howpublished={\url{https://docs.anthropic.com/en/claude-code}},
    note={IDE-integrated autonomous coding agent built on Claude large language models}
}

@misc{openai2021codexweb,
    author = {{OpenAI}},
    title = {Introducing Codex},
    year = {2021},
    howpublished = {\url{https://openai.com/research/codex}},
    note = {Accessed: 2026-06-25}
}

@misc{mitre2026cwe416,
    author = {{The MITRE Corporation}},
    title = {CWE-416: Use After Free},
    year = {2026},
    howpublished = {\url{https://cwe.mitre.org/data/definitions/416.html}},
    note = {Accessed: 2026-06-25, Common Weakness Enumeration (CWE)}
}

@misc{tree-sitter,
    author = {{Max Brunsfeld et al.}},
    title = {Tree-sitter: An incremental parsing system for programming tools},
    year = {2024},
    howpublished = {\url{https://tree-sitter.github.io/tree-sitter/}},
    note = {Accessed: 2026-06-26}
}

@article{DBLP:journals/corr/abs-2509-22908,
  author       = {Sergiu Bursuc and
                  Theodore Ehrenborg and
                  Shaowei Lin and
                  Lacramioara Astefanoaei and
                  Ionel Emilian Chiosa and
                  Jure Kukovec and
                  Alok Singh and
                  Oliver Butterley and
                  Adem Bizid and
                  Quinn Dougherty and
                  Miranda Zhao and
                  Max Tan and
                  Max Tegmark},
  title        = {A benchmark for vericoding: formally verified program synthesis},
  journal      = {CoRR},
  volume       = {abs/2509.22908},
  year         = {2025},
  url          = {https://doi.org/10.48550/arXiv.2509.22908},
  doi          = {10.48550/ARXIV.2509.22908},
  eprinttype   = {arXiv},
  eprint       = {2509.22908},
  bibsource    = {dblp computer science bibliography, https://dblp.org}
}

@inproceedings{DBLP:conf/acl/XinXYCWXSZD25,
  author       = {Ran Xin and
                  Chenguang Xi and
                  Jie Yang and
                  Feng Chen and
                  Hang Wu and
                  Xia Xiao and
                  Yifan Sun and
                  Shen Zheng and
                  Ming Ding},
  editor       = {Wanxiang Che and
                  Joyce Nabende and
                  Ekaterina Shutova and
                  Mohammad Taher Pilehvar},
  title        = {BFS-Prover: Scalable Best-First Tree Search for LLM-based Automatic
                  Theorem Proving},
  booktitle    = {Proceedings of the 63rd Annual Meeting of the Association for Computational
                  Linguistics (Volume 1: Long Papers), {ACL} 2025, Vienna, Austria,
                  July 27 - August 1, 2025},
  pages        = {32588--32599},
  publisher    = {Association for Computational Linguistics},
  year         = {2025},
  url          = {https://doi.org/10.18653/v1/2025.acl-long.1565},
  doi          = {10.18653/V1/2025.ACL-LONG.1565},
  bibsource    = {dblp computer science bibliography, https://dblp.org}
}

@article{DBLP:journals/corr/abs-2504-21801,
  author       = {Z. Z. Ren and
                  Zhihong Shao and
                  Junxiao Song and
                  Huajian Xin and
                  Haocheng Wang and
                  Wanjia Zhao and
                  Liyue Zhang and
                  Zhe Fu and
                  Qihao Zhu and
                  Dejian Yang and
                  Z. F. Wu and
                  Zhibin Gou and
                  Shirong Ma and
                  Hongxuan Tang and
                  Yuxuan Liu and
                  Wenjun Gao and
                  Daya Guo and
                  Chong Ruan},
  title        = {DeepSeek-Prover-V2: Advancing Formal Mathematical Reasoning via Reinforcement
                  Learning for Subgoal Decomposition},
  journal      = {CoRR},
  volume       = {abs/2504.21801},
  year         = {2025},
  url          = {https://doi.org/10.48550/arXiv.2504.21801},
  doi          = {10.48550/ARXIV.2504.21801},
  eprinttype   = {arXiv},
  eprint       = {2504.21801},
  bibsource    = {dblp computer science bibliography, https://dblp.org}
}

\end{document}